\documentclass[aps,prd,12pt,reprint,nofootinbib,superscriptaddress]{revtex4-2}
\usepackage{bm}
\usepackage{amsmath}
\usepackage{ulem}
\usepackage{changes}
\usepackage{latexsym}
\usepackage{amsfonts}
\usepackage{graphicx}
\usepackage{mathrsfs}
\usepackage{CJKutf8}
\usepackage{subfigure}
\usepackage{float}
\usepackage{multirow}
\usepackage{color}
\usepackage{comment}
\usepackage{hyperref}
\hypersetup{
	colorlinks=true,
	linkcolor=red,
	citecolor=blue,
}

\begin{document}

\begin{CJK*}{UTF8}{gbsn} 

\title{Intermediate mass-ratio inspirals with dark matter minispikes}

\author{Ning Dai (戴宁)}
\email{daining@hust.edu.cn}
\affiliation{School of Physics, Huazhong University of Science and Technology, Wuhan, Hubei
430074, China}

\author{Yungui Gong (龚云贵)}
\email{Corresponding author. yggong@hust.edu.cn}
\affiliation{School of Physics, Huazhong University of Science and Technology, Wuhan, Hubei
430074, China}

\author{Tong Jiang (江通)}
\email{jiangtong@hust.edu.cn}
\affiliation{School of Physics, Huazhong University of Science and Technology, Wuhan, Hubei
430074, China}

\author{Dicong Liang (梁迪聪)}
\email{dcliang@pku.edu.cn}
\affiliation{Kavli Institute for Astronomy and Astrophysics, Peking University, Beijing
100871, China}

\begin{abstract}
The dark matter (DM) distributed around an intermediate massive black hole (IMBH) forms an overdensity region called DM minispike.
We consider the binary system which consists of an IMBH with DM minispike and a small black hole inspiralling around the IMBH in eccentric orbits.
The factors which affect the evolution of the orbit include the gravity of the system, the dynamical friction and accretion of the small black hole caused by the DM minispike, and the radiation reaction of gravitational waves (GWs).
Using the method of osculating orbit, we find that when the semi-latus rectum $p\ll 10^5 R_s$ ($R_s$ is the Schwarzschild radius of the IMBH) the dominated factors are the dynamical friction and accretion from the DM minispike, and the radiation reaction.
When $p\gg 10^5 R_s$,
the gravity from the DM minispike dominates the orbital evolution.
The existence of DM minispike leads to the deviation from the Keplerian orbit,
such as extra orbital precession,
henceforth extra phase shift in the GW waveform.
By calculating the signal-to-noise ratio for GWs with and without DM minispikes and the mismatch between them, 
we show that the effect of the DM minispike in GW waveforms can potentially be detected by future space-based GW detectors such as LISA, Taiji, and Tianqin.
\end{abstract}

\maketitle

\end{CJK*}

\section{introduction}

Although there is a large amount of observational evidence from different scales on the existence of dark matter (DM) which accounts for 26\% of the total mass of the Universe \cite{Bertone:2004pz,Clowe:2006eq,Planck:2018vyg},
we still know nothing about the nature and origin of DM.
The study of DM is of great importance for understanding the formation and evolution of the Universe and finding possible breakthrough in fundamental physics \cite{Bertone:2004pz,Bertone:2018krk}.

Because of the extremely strong gravity around black holes (BHs),
there might exist DM halos around them.
It was pointed out by Navarro, Frenk, and White (NFW) that all equilibrium density profiles of DM halos have the same shape, which is called NFW profile \cite{Navarro:1996gj}.
Then Gondolo and Silk suggested that the adiabatic growth of supermassive black holes (SMBHs) with masses $10^6\sim 10^9\ M_{\odot }$ would generate overdensity DM regions around them, called DM spikes \cite{Gondolo:1999ef}.
However, for SMBHs, DM spike could be disrupted and form a light density region as a result of galaxy merger or other astronomical activities \cite{Merritt:2002vj,Ullio:2001fb,Bertone:2005xv,Vasiliev:2008uz}.
With the role of these effects in doubt \cite{Fields:2014pia,Shelton:2015aqa},
it is more likely that DM spike exists around the intermediate-massive black holes (IMBHs) with masses $10^2\sim 10^5 M_{\odot }$, which is called minispike \cite{Zhao:2005zr,Bertone:2005xz}.
The gravity of DM spike could affect the orbit of the small object moving around the central BH \cite{Sadeghian:2013laa,Ferrer:2017xwm}.
Optical observation of the orbital motion of the small object can be used to test the existence of DM spike indirectly and constrain the density profile of the spike \cite{Bertone:2004pz,Takamori:2020ntj}.
Due to the effect of the DM minispike on the orbital motion of binaries,
the observations of gravitational waves (GWs) emitted by these binaries can also be used to detect DM minispikes \cite{Eda:2013gg,Eda:2014kra,Barausse:2014tra,Yue:2017iwc,Yue:2018vtk,Hannuksela:2019vip,Cardoso:2019rou,Cardoso:2021wlq}.

Since the detection of the first binary BH and the first binary neutron star mergers \cite{Abbott:2016blz,TheLIGOScientific:2017qsa},
there have been tens of GW events detected which
opened a new window for the test of gravity in the strong field and nonlinear regions
\cite{Monitor:2017mdv,LIGOScientific:2018mvr,LIGOScientific:2020ibl,LIGOScientific:2021usb,LIGOScientific:2021djp}.
In particular, the 90\% credible intervals for the mass of the remnant BH in GW190521 are $163.9^{+39.2}_{-23.5}M_{\odot}$ \cite{Abbott:2020tfl}.
This is the most massive merger remnant observed so far and it provides a direct observation of the formation of an IMBH \cite{LIGOScientific:2020ufj}.
Additionally, there are four more GW events-- GW190519, GW190602, GW190706, and GW190929--
with the mass of the remnant BH heavier than $100M_{\odot}$.
IMBHs may come from primordial BHs formed as a result of gravitational collapse in overdense regions with their density contrasts at the horizon reentry during radiation domination exceeding the threshold value \cite{Carr:1974nx,Hawking:1971ei}. 
Astrophysically, IMBHs may form from the evolution of nearly zero metallicity Population III stars \cite{Madau:2001sc}, the mass segregation, runaway collision and merging in dense, young clusters \cite{1969ApJ...158L.139S,Rasio:2003sz,AtakanGurkan:2003hm,PortegiesZwart:2004ggg,Mapelli:2016vca}, the gas accretion on to stellar-mass BHs \cite{Miller:2001ez,Leigh:2013tqa}, 
binary dynamical interaction, and mass transfer in binaries in dense star clusters \cite{2015MNRAS.454.3150G}.
Stars are removed from the cluster by tidal stripping and ejection and the IMBH is released \cite{Miller:2001ez,Miller:2004va}. 
It was suggested IMBHs with the mass of $\sim 10^3$ may exist in some tens of per cent of current globulars \cite{Miller:2001ez}, 
and even hundreds of IMBHs are present in the Galactic bulge and halo \cite{Islam:2002gg,Rashkov:2013uua}.
However, there is a great uncertainty about the population of IMBHs and observational evidence of IBMHs remains in dispute.
For a review on the formation and evidence of IMBHs, please see Ref. \cite{Miller:2003sc,Feng:2011pc,Kaaret:2017tcn}.

As an IMBH sinks to the center, it is possible for the IMBH to capture a companion and the binary was hardened by repeated interactions.
A small compact object captured into the inspiral orbits around an IMBH/SMBH forms an intermediate-mass-ratio ($10^2\sim 10^4$) inspiral (IMRI) or an extreme-mass-ratio ($10^4\sim 10^6$) inspiral (EMRI) system.
For EMRI/IMRI, the small compact object spends the last few years inspiralling 
deep inside the strong gravitational field around the massive BH (MBH) with a highly relativistic speed.
The emitted GWs from IMRI/EMRI encode rich information about the spacetime geometry around the MBH and the environment of the host galaxy,
so they can be used to confirm whether the MBH is a Kerr BH predicated by GR.
Therefore, the study of IMRIs/EMRIs cannot only tell us information about the dynamics of large-mass-ratio binaries and the property and growth of BHs,
but also sheds light on fundamental physics such as dark matter, dark energy, and quantum gravity \cite{Amaro-Seoane:2007osp,eLISA:2013xep}.
As long-duration sources of GWs, 
there are thousands of GW cycles in the detector band 
of space-based GW detectors such as LISA \cite{Audley:2017drz}, Taiji \cite{Hu:2017mde}, and TianQin \cite{Luo:2015ght,Gong:2021gvw}.
The event rate depends on a number of factors,
such as the fraction of star clusters with a MBH,
the mass distribution of BHs and the mechanism for the formation of MBHs, etc. \cite{Miller:2003sc}.
It was estimated that LISA could detect IMRIs with an event rate $\sim 3-10$ Gpc$^{-3}$ yr$^{-1}$ \cite{Fragione:2017blf}, or 10 IMRIs consisting of BHs with $10^3\,M_\odot$ and $10\,M_\odot$ at any given time \cite{Miller:2001ez},
or a few IMRIs/EMRIs consisting of an IMBH and a SMBH per year \cite{Miller:2004va}.

When a small object moves around an IMBH with DM minispike, it is affected by the gravity of the central BH and the DM minispike \cite{Eda:2013gg,Eda:2014kra,Barausse:2014tra,Yue:2017iwc,Yue:2018vtk,Hannuksela:2019vip,Cardoso:2019rou,Cardoso:2021wlq}.
Besides, the small object is driven by the gravitational drag (dynamical friction) (DF) of the DM minispike while moving through the DM minispike \cite{Chandrasekhar:1943ys,Ostriker:1998fa,Kim:2007zb}.
Considering the effects of gravity and the DF of DM minispike and GW reaction,
analytical GW waveforms were derived in \cite{Eda:2014kra} for IMRIs in quasi-circular orbits to Newtonian order
by assuming a single power-law model for the DM minispikes,
and the power-law index $\alpha$ can be determined to $10\%$ accuracy for $\alpha\sim 1.7$ with LISA
for IMRIs composed of an IMBH with mass $10^3\, M_{\odot}$ and a compact object with mass $1\, M_{\odot}$ in quasi-circular orbits \cite{Eda:2014kra}.
Due to its gravitational interaction with the binary, 
the DM minispike surrounding IMBH could evolve \cite{Kavanagh:2020cfn}.
The DM density profile is not static because there is an efficient transfer of energy
from the binary to the DM spike and the energy dissipated by the compact object through DF can be much larger than the gravitational binding energy in the DM distribution,
so the dephasing of the gravitational waveform induced by the DF was overestimated with the assumption of a fixed DM density profile, 
but it is still potentially detectable with LISA even if the evolution of the DM minispike is taken into account \cite{Kavanagh:2020cfn}.
If the small object is a BH, it also accretes the medium surrounding it \cite{Bondi:1944jm,Shapiro:1983du}.
Different types of accretion and DF of the DM minispikes have different effects on the evolution of IMRIs \cite{Macedo:2013qea}.
Including the effect of accretion in addition to the effects of gravity, DF and GW reaction,
the authors of \cite{Yue:2017iwc} calculated the time and phase differences caused by DM minispikes using the same method used in \cite{Eda:2014kra}
and they found that the inspiral time is reduced dramatically for smaller IMBHs and larger $\alpha$ and the time difference is detectable with LISA.
They \cite{Yue:2017iwc} also compared the contribution to the phase difference with and without the accretion effect
and they found that the contribution to the phase difference is dominant by the DF
and the accumulated phase shift caused by the accretion effect only can be detected with LISA, Taiji and TianQin.
Because DF and accretion cause the orbit of the binary to decay faster, 
the existence of the DM minispikes could be an efficient catalyst for the merger of IMRIs \cite{Yue:2018vtk}.

The eccentricity of a binary may not be small at merger \cite{Kozai:1962zz,Heggie:1975tg,Wen:2002km,Miller:2002pg,Hoang:2019kye},
so it is necessary to consider eccentric IMRIs to understand astrophysical formation channels of binaries and the properties of DM minispikes \cite{Cardoso:2020iji}.
The orbital eccentricity could increase under the influence of dynamical friction (DF) of the surrounding medium such as DM 
and decrease by GW reaction \cite{Yue:2019ozq,Cardoso:2020iji}.
In this paper, we study the effects of DM minispike on the orbital motion and GW waveforms by using the method of osculating orbital perturbation \cite{Gerhard:2005mcm,Poisson:2018gn}.
We consider IMRIs with DM minispikes in eccentric orbits to discuss the effects of the gravity of the central IMBH and DM minispike, the DF of DM minispike, the accretion of the small BH and the radiation reaction of GWs, separately and concurrently.
The paper is organized as follows.
In Sec. \ref{DM-influence}, we discuss each of the above effects on the orbital motion.
The combined effects on the orbital motion and GW waveforms are discussed in Sec. \ref{Combined Effect}.
We also calculate the signal-to-noise ratio (SNR) for and the mismatch between GWs from IMRIs in eccentric orbits with and without DM minispikes in Sec. \ref{Combined Effect}.
We draw the conclusion in Sec. \ref{conclusions}.
The details of the method of osculating orbital perturbation is presented in Appendix \ref{APPENDIX}.
We present the results of parameter estimation for IMRIs in circular orbits with the method of Fisher information matrix (FIM) in Appendix \ref{APPENDIX2}.

\section{The effects of the DM minispike}
\label{DM-influence}

In this section, we consider IMRIs consisting of an IMBH surrounded by a DM minispike and a stellar mass BH inspiralling around the IMBH.
The motion of the IMRI is affected by several dynamical factors, 
such as the gravity of both the IMBH and the DM minispike, the DF, 
the accretion of the small BH, and the radiation reaction of GWs.
We discuss the effect of each factor in this section.

\subsection{Gravity of the DM minispike}
\label{DMdist}

We choose the mass of the IMBH as $M=10^3 M_{\odot}$ and the mass of the small BH as $\mu=10M_{\odot}$.
For this IMRI, the reduced mass $\epsilon$ and total mass $m$ are approximately equal to
$\mu$ and $M$, respectively, $\epsilon=M\mu/(M+\mu)\simeq \mu$ and $m=M+\mu\simeq M$.
Following \cite{Eda:2013gg,Eda:2014kra}, we adopt the distribution of DM
\begin{eqnarray}\label{p-spike}
\rho_{\text{DM}} (r)=
    \begin{cases}
\rho_{\text{sp}}\left(\frac{r_{\text{sp}}}{r}\right)^{\alpha},  & r_{\text{min}}\leq r\leq r_{\text{sp}}, \\
0,  & r\leq r_{\text{min}},
    \end{cases}
\end{eqnarray}
where $r$ is the distance from the test point to the central IMBH,
$r_{\text{sp}}$ is used to characterize the range of the DM minispike,
$\rho_{\text{sp}}$ is the DM density at the distance $r_{\text{sp}}$,
and $r_{min}$ is chosen to be the innermost stable circular orbit (ISCO) of the central IMBH,
$r_\text{min}=r_\text{ISCO}=3R_\text{s}=6G M/{c^2}$.
For the central IMBH with the mass $10^3 M_{\odot}$,
we have $r_{\text{sp}}=0.54 \ \text{pc}$ and $\rho_{\text{sp}}=226M_{\odot}/\text{pc}^3$ \cite{Eda:2013gg,Eda:2014kra}.
The power index $\alpha=(9-2\alpha_\text{ini})/(4-\alpha_\text{ini})$ with the initial profile parameter $\alpha_\text{ini}$ describing the final profile of DM halo,
which depends on the formation history of the central IMBH.
Take the NFW case as an example, the initial profile parameter is $\alpha_\text{ini}=1$,
henceforth $\alpha=7/3$ \cite{Navarro:1996gj}.
For the DM spike, $0\leq \alpha_\text{ini}\leq 2$, so $2.25\leq\alpha\leq 2.5$ \cite{Gondolo:1999ef}.
For the DM region distributed around IMBH, the range of $\alpha$ maybe wider \cite{Eda:2013gg,Eda:2014kra}.
In this paper, we adopt $2.25\leq\alpha\leq 2.5$.
The mass of the DM minispike within $r$ is
\begin{eqnarray}\label{M-spike}
    M_\text{DM}=
\begin{cases}
    \frac{4\pi\rho_{\text{sp}}r_{\text{sp}}^{\alpha}}{3-\alpha}\left(r^{3-\alpha}-r_{\text{min}}^{3-\alpha}\right), &r_{\text{min}}\leq r\leq r_{\text{sp}}, \\
    0, &r\leq r_{\text{min}}.
\end{cases}
\end{eqnarray}
With Eq. \eqref{M-spike}, the acceleration of the small BH is
\begin{equation}\label{a-m+dm}
    \bm{a}_{\text{G}}=-\frac{G M_{\text{eff}}}{r^2}\bm{n}-\frac{G F}{r^{\alpha-1}}\bm{n},
\end{equation}
where $M_{\text{eff}}=M-4\pi\rho_{\text{sp}}r_{\text{sp}}^{\alpha}{r_{\text{min}}}^{3-\alpha}/(3-\alpha)$,
$F=4\pi\rho_{\text{sp}}r_{\text{sp}}^{\alpha}/(3-\alpha)$ and $\bm{n}$ is the unit vector pointing from the central IMBH to the small BH.
The first term in Eq. \eqref{a-m+dm} mainly comes from the gravitational interaction of IMBH and the second term is from the DM minispike.
When $\alpha-1\neq 2$, the second term is not in the form of inverse square law,
so the orbit of the small BH is no longer Keplerian.
We take the second term as perturbation and use the osculating orbit method to discuss the deviation from the Keplerian orbit.
Comparing Eq. \eqref{a-m+dm} with Eq. \eqref{a-k} in Appendix \ref{APPENDIX},
we have
\begin{equation}\label{f-m+dm}
    \bm{f}_{\text{G}}=-\frac{G F}{r^{\alpha-1}}\bm{n}=\mathcal{R}_\text{G} \bm{n} .
\end{equation}

Substituting Eq. \eqref{f-m+dm} into Eqs. \eqref{pk-pf}, \eqref{pk-ef}, \eqref{pk-wf} and \eqref{pk-tf}, the osculating equations can be written as
\begin{align}
    \label{dm-pf}
    \frac{d p}{d \phi}=& 0 ,\\
    \label{dm-ef}
    \frac{d e}{d \phi}=& -\frac{p^{3-\alpha}F}{M_{\text{eff}}}\frac{\sin{\phi}}{(1+e\cos\phi)^{3-\alpha}} ,\\
    \label{dm-wf}
    \frac{d \omega}{d \phi}= &\frac{p^{3-\alpha}F}{M_{\text{eff}}}\frac{\cos{\phi}}{e(1+e\cos\phi)^{3-\alpha}} ,\\
    \label{dm-tf}
    \frac{d t}{d \phi}=& \sqrt{\frac{p^3}{G M_{\text{eff}}}}\frac{1}{(1+e\cos\phi)^{2}}\nonumber\\
    &\times\left[1+\frac{p^{3-\alpha}F\cos{\phi}}{M_{\text{eff}}\,e(1+e\cos\phi)^{3-\alpha}}\right],
\end{align}
where $\phi$ is the true anomaly angle,
$p$ is the semi-latus rectum,
$e$ is the orbital eccentricity, $\omega$ is the longitude of pericenter,
${d \omega}/{d \phi}$ describes the pericenter precession of the orbit.
Comparing with the Keplerian motion, the terms including $p^{3-\alpha}$ in the above equations are the correction from the DM minispike.

\begin{figure}[tbp]
    \centering
     \includegraphics[width=0.46
     \textwidth,origin=c]{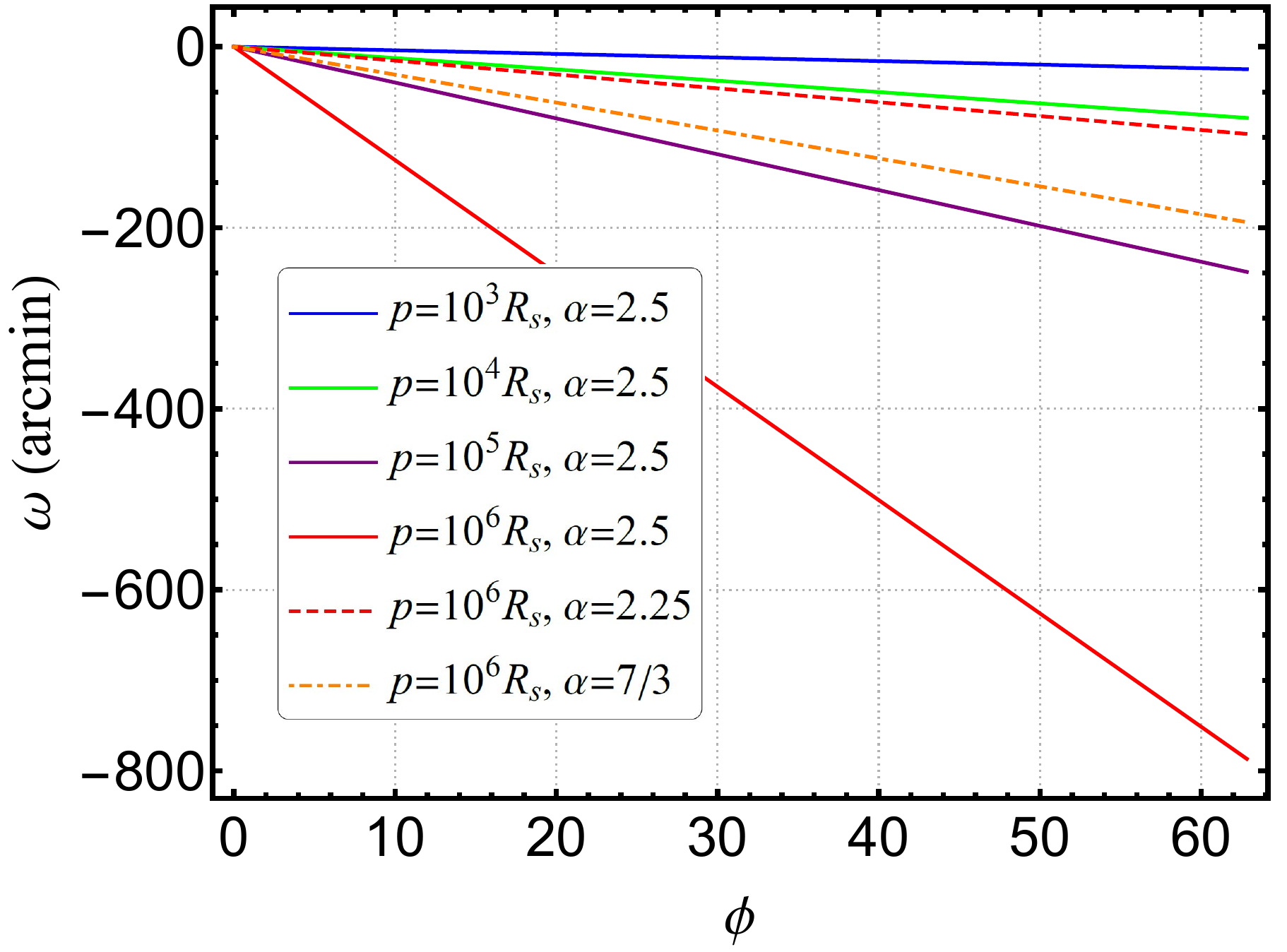}
    \caption{
     The accumulated $\omega$ versus $\phi$ for different $p$ and $\alpha$ under the influence of the gravity from the DM minispike.
     The eccentricity $e$ is $0.6$, and the values of $\alpha$ are chosen as $2.25$, $7/3$ and $2.5$.
     The semi-latus rectum $p$ are chosen as $10^3 R_s$, $10^4 R_s$, $10^5 R_s$ and $10^6 R_s$.}
\label{fig:2.1}
\end{figure}

Combining Eqs. \eqref{dm-ef}, \eqref{dm-wf}, and \eqref{accmulated},
we obtain the accumulated changes of $e$ and $\omega$ over one period,
\begin{align}
    \label{de-dm}
    \Delta e &=0 ,\\
    \label{wa-dm}
    \Delta\omega_{\text{DM}}&=\frac{p^{3-\alpha}F}{M_{\text{eff}}}W_{\text{DM}}(e),
\end{align}
where $W_{\text{DM}}(e)=\int_{0}^{2\pi} \cos\phi(1+e\cos\phi)^{\alpha-3}e^{-1}  \,d\phi$,
and the subscript DM means the gravitational effect of the DM minispike.
Note that $W_{\text{DM}}$ is always less than zero when $0<e<1$ and $1<\alpha<3$.
If $\rho_\text{sp}=0$, i.e. there is no DM, then the right-hand side of Eq. \eqref{wa-dm} becomes zero and  $\Delta\omega_{\text{DM}}$ is zero.
From Eq. \eqref{de-dm}, we see that the accumulated change of $e$ in one period is zero.
The accumulated change of $\omega$ will cause an additional orbital precession, as seen from  Eq. \eqref{wa-dm}.
We plot the accumulated $\omega$ versus $\phi$ for different $p$ and $\alpha$ in Fig. \ref{fig:2.1}.
As shown in Fig. \ref{fig:2.1}, $\omega$ does not evolve much for $p=10^3R_\text{s}$ regardless the value of $\alpha$,
but its change is not small for $p=10^6 R_\text{s}$.
The larger value of $\alpha$, the larger amplitude of the precession.
These results can be easily understood because
the total mass of the DM minispike within the region $p\leq 10^3 R_s$ is small, so the gravitational effect of DM minispike is negligible.
At large orbital distance $p=10^6R_\text{s}$, the gravity caused by the DM minispike can not be ignored, so the effect of the DM minispike on the orbital motion becomes important.

However, the orbit also experiences the relativistic precession caused by the higher-order effect of gravitational interaction \cite{Will:2007pp,Will:2018bme}.
Using the post-Newtonian results \cite{Poisson:2018gn},
the change of the relativistic precession with DM minispike over one orbital period is
\begin{align}
\label{wa-pn}
        \Delta\omega_{\text{rp}} \simeq \frac{6\pi G M_{\text{eff}}}{c^2 p}+\frac{G F}{c^2}p^{2-\alpha}W_{\text{rp}}(e),
\end{align}
where
\begin{align}
        W_{\text{rp}}(e)=\int_{0}^{2\pi}\frac{(3-e^2)\cos\phi-5e\cos{2\phi}+3e}{e(1+e\cos\phi)^{3-\alpha}}\,d\phi. \nonumber
\end{align}
The subscript 'rp' means the relativistic precession.
$W_{\text{rp}}$ is greater than zero when $0<e<1$ and $1<\alpha<3$.

\begin{figure}[H]
    \centering
    \includegraphics[width=.45\textwidth]{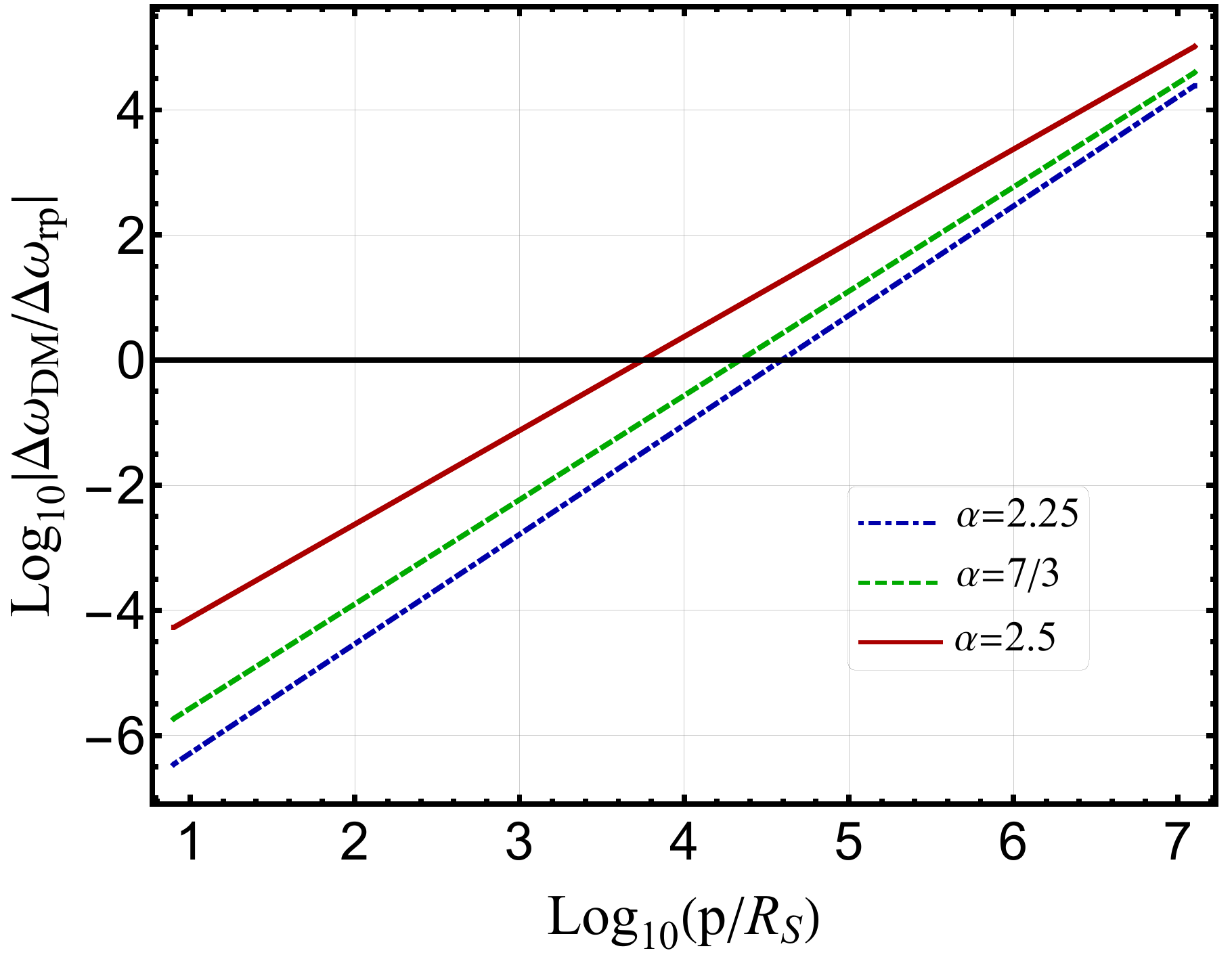}
    \caption{
    The ratio $\Delta\omega_{\text{DM}}/\Delta\omega_{\text{rp}}$ versus the semi-latus rectum $p$ in the range $3R_\text{s}$ to $10^8R_\text{s}$ for different values of $\alpha$. We take the orbital eccentricity $e=0.6$.
    }
    \label{fig:2.2}
\end{figure}

To compare the effects of gravitational interaction and relativistic precession at different orbital distance,
we show $\Delta\omega_{\text{DM}}/\Delta\omega_{\text{rp}}$ with respect to $p$ for different values of $\alpha$ in Fig. \ref{fig:2.2}.
We see that at small orbital distance $p\ll  10^5 R_\text{s}$,
$\Delta\omega_{\text{rp}}$ is greater than $\Delta\omega_{\text{DM}}$, and it may even be $\sim 4-5$ orders of magnitude greater,
so the effect of the relativistic precession dominates in this region.
At large orbital distance $p\gg  10^5 R_\text{s}$,  
$\Delta\omega_{\text{DM}}$ is much larger.
For example, $\Delta\omega_{\text{DM}}/\Delta\omega_{\text{rp}}\sim 10^4$ when $p=10^7 R_\text{s}$. 
Therefore, the effect of the gravity of the DM minispike dominates at large orbital distance where $p\gg 10^5 R_\text{s}$.

\subsection{Dynamical friction and accretion}
\label{DFA}

Chandrasekhar suggested that moving  objects may be dragged by the gravity of the interstellar medium particles,
this is called DF \cite{Chandrasekhar:1943ys}.
The property of DF depends on the velocity of the moving object,
the density and the sound speed of the medium \cite{Ostriker:1998fa,Kim:2007zb}.

While moving through the DM minispike around the central IMBH, the small BH is dragged by the DF of the DM minispike.
Without loss of generality, we discuss cases in supersonic regime that the DF can be described as \cite{Cardoso:2020iji}
\begin{equation} \label{f-df}
   \bm{f}_{\text{DF}}=-\frac{4\pi G^2 {\mu}^2 \rho_{\text{DM}} I_{v}}{v^3} \bm{v},
\end{equation}
where $\bm{v}$ is the velocity of the small BH,
$I_v$ is the Coulomb logarithm which depends on $v$ and the sound speed of the DM minispike. In this paper, we adopt $I_v=3$ \cite{Eda:2014kra}.

Substituting Eq. \eqref{f-df} into Eqs. \eqref{pk-pf}, \eqref{pk-ef}, and \eqref{pk-wf} and averaging the result
(the orbital average of a physical variable is defined in Eq. \eqref{average}), we have
\begin{align}
        \label{pk-df-pf}
        \left\langle\frac{d p}{d \phi}\right\rangle_{\text{DF}}&=-\frac{4\mu\rho_{\text{sp}} r_{\text{sp}}^{\alpha}I_v}{M^2}p^{4-\alpha} g(e),\\
        \label{pk-df-ef}
        \left\langle\frac{d e}{d \phi}\right\rangle_{\text{DF}}&=-\frac{4\mu\rho_{\text{sp}} r_{\text{sp}}^{\alpha}I_v}{M^2}p^{3-\alpha} f(e),\\
        \label{pk-df-wf}
        \left\langle\frac{d \omega}{d \phi}\right\rangle_{\text{DF}}&=0,
\end{align}
where
\begin{align}
        \label{pk1-df-pf}
        g(e)=\int_{0}^{2\pi} \frac{d{\phi}}{(1+2e\cos{\phi}+e^2)^{3/2}(1+e\cos{\phi})^{2-\alpha}} ,\\
        \label{pk1-df-ef}
        f(e)=\int_{0}^{2\pi} \frac{(\cos{\phi}+e) d{\phi} }{(1+2e\cos{\phi}+e^2)^{3/2}(1+e\cos{\phi})^{2-\alpha}}  ,
\end{align}
and the subscript 'DF' means that it is due to the effect of DF.
It is obvious that $g(e)$ is always greater than 0.
When $0<e<1$ and $1<\alpha<3$, $f(e)$ is less than zero.
Combining Eqs. \eqref{f-df} and \eqref{pk-tf}, we obtain
\begin{align}\label{pk-df-tf}
\begin{split}
       \left(\frac{d t}{d \phi}\right)_{\text{DF}}= &
       \frac{\sqrt{p^3/(GM)}}{(1+e\cos\phi)^2} \times \bigg(1- \\
        &\frac{8\pi \,\mu\, \rho_{\text{DM}}\,r_{\text{sp}}^{\alpha}\,I_v\, p^{3-\alpha}\,\sin{\phi}}{m^2\,e\,(1+2e\cos{\phi}+e^2)^{3/2}(1+e\cos{\phi})^{2-\alpha}}\bigg).
\end{split}
\end{align}
In the above equation, the second term in the brackets is the correction from DF.
Take the orbital average of Eq. \eqref{pk-df-tf}, we get
\begin{equation}\label{pk-df1-tf}
        \left\langle\frac{d t}{d \phi}\right\rangle_{\text{DF}}= \sqrt{\frac{p^3}{G M}}(1-e^2)^{-3/2},
\end{equation}
which is the same as Keplerian motion.

\begin{figure}[H]
    \centering
    \includegraphics[width=.46\textwidth]{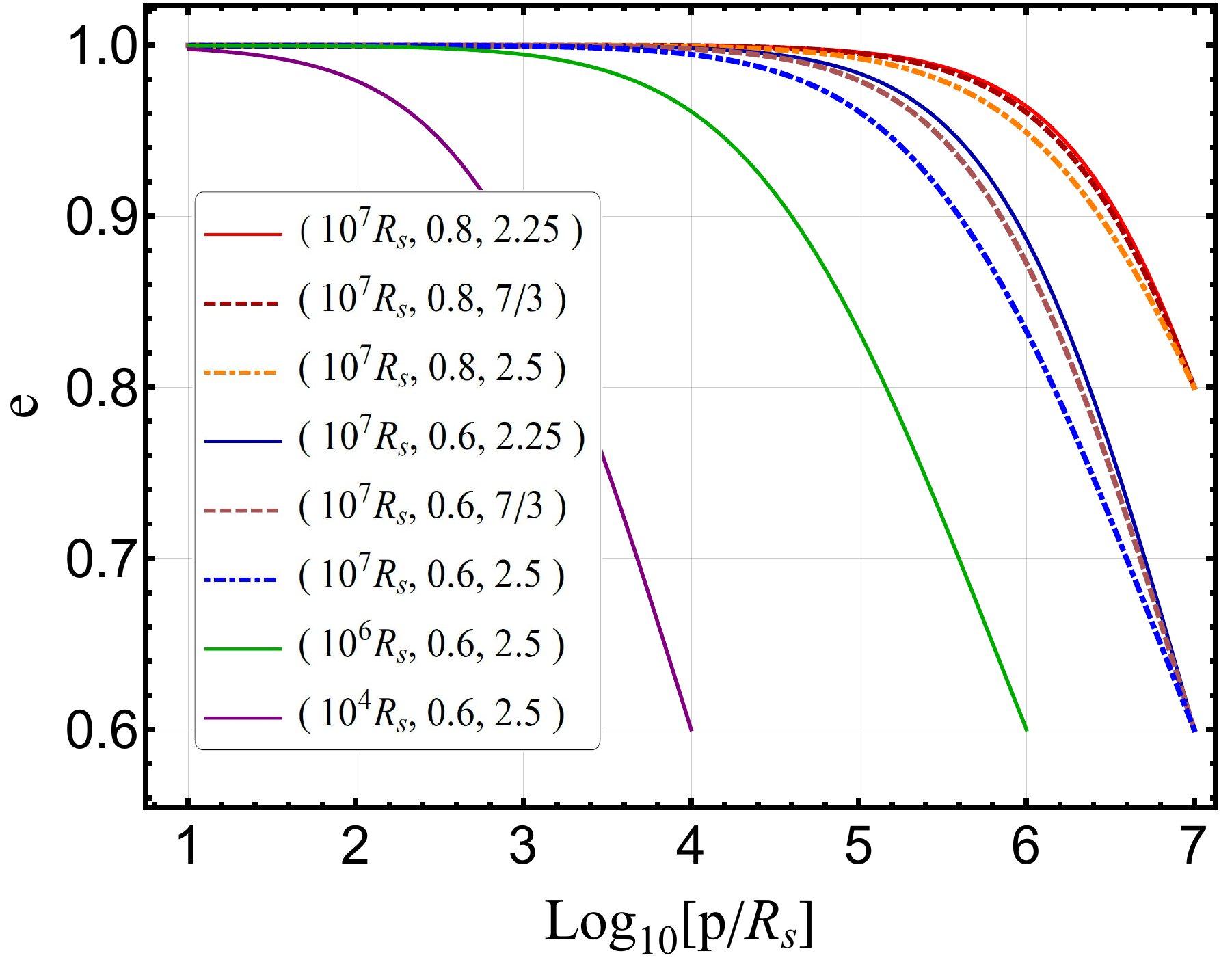}
    \caption{
        The eccentricity $e$ versus the semi-latus rectum $p$ under the influence of DF.
        We take the initial eccentricity $e_0$ as $0.6$ and $0.8$,
        the initial semi-latus rectum $p_0$ as $10^4 R_\text{s}$, $10^6 R_\text{s}$, and $10^7 R_\text{s}$,
        and the values of $\alpha$ as $2.25$, $7/3$ and $2.5$.
        In the legends, $(10^7R_s,0.8,2.25)$ meas that $p_0=10^7R_s$, $e_0=0.8$, and $\alpha=2.25$. 
        }
    \label{fig:2.3}
\end{figure}

\begin{figure}[H]
    \centering
    \includegraphics[width=.46\textwidth]{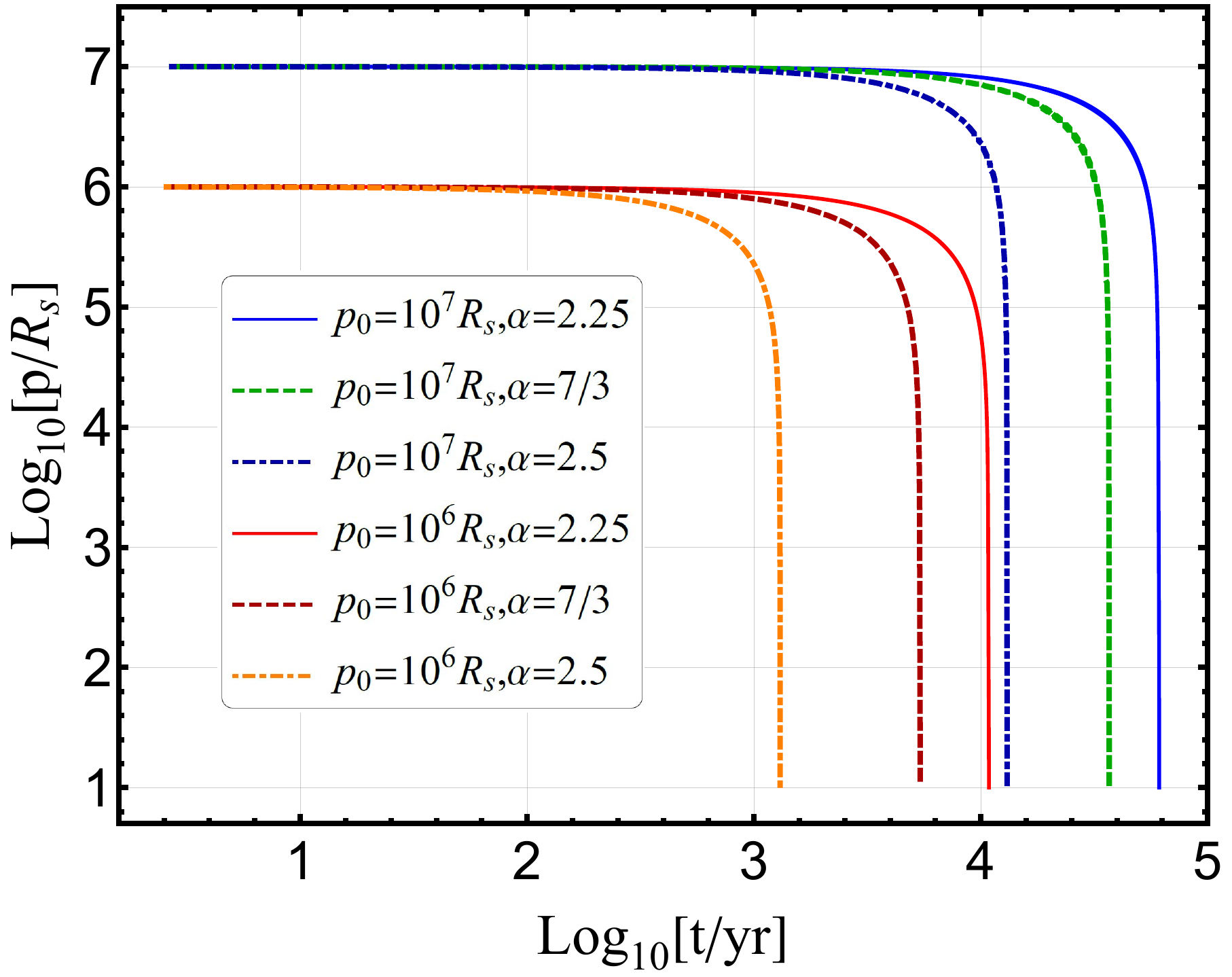}
    \caption{
        The evolution of $p$ under the influence of DF.
        We take the initial eccentricity $e_0$ as $0.6$,
        the initial semi-latus rectum $p_0$ as $10^6 R_\text{s}$, and $10^7 R_\text{s}$, and
        the values of $\alpha$ as $2.25$, $7/3$, and $2.5$.
        }
    \label{fig:2.4}
\end{figure}

From Eqs. \eqref{pk-df-pf}, \eqref{pk-df-ef}, and \eqref{pk-df-wf},
we see that DF acts as a dissipated force.
Under the influence of DF, the orbital radius of the system decreases and the eccentricity increases.
However, DF does not affect the orbital precession.
We plot the evolution of $e$ versus $p$ for different $p_0$, $e_0$, and $\alpha$ in Fig. \ref{fig:2.3},
and show the changes of $p$ with respect to $t$ for different values of $p_0$ and $\alpha$ in Fig. \ref{fig:2.4}.

As shown in Fig. \ref{fig:2.3}, the eccentricity $e$ increases as $p$ decreases under the influence of DF,
so all orbits will evolve to $e\rightarrow 1$ as $p\rightarrow 0$, i.e. head-to-head collision of the binary if only DF is considered and post-Newtonian result is valid.
From Fig. \ref{fig:2.4} we can see the orbit decays with time.
Before the orbit shrink in a short time, it evolves slowly for a long time, even more than thousands of years.
The value of $\alpha$ is greater, the orbit starts the fast shrink  earlier.

Now we turn to the discussion of accretion.
The accretion of the small BH we considered is characterized as Bondi-Hoyle accretion \cite{Bondi:1944jm,Edgar:2004mk}.
We assume that the radius of the small BH is greater than the mean free path of DM particles, so
the mass flux at the horizon of the small BH is \cite{Macedo:2013qea,Mach:2021zqe}
\begin{equation} \label{ut-dm}
    \dot{\mu}=4\pi G^2 \lambda\frac{{\mu}^2 \rho_{\text{DM}}}{(v^2+c_s^2)^{3/2}} ,
\end{equation}
where $\lambda$ is of order one and depends on the DM medium,
and $c_s$ is the sound speed of the DM medium.
For simplicity, we assume $v\gg c_s$ and $\lambda=1$ in this paper.

Considering the influence of the accretion only, the orbital equation of motion is
\begin{equation} \label{eom-a-dm}
    \mu\dot{\bm{v}}+\dot{\mu}\bm{v} = -\frac{G\mu M}{r^3}\bm{n}.
\end{equation}
The accretion term $\dot{\mu}\bm{v}$  can be thought as a perturbation force,
\begin{equation}\label{f-aa}
    \bm{f}_{\text{a}}\simeq -\frac{4\pi G^2 \mu^2 \rho_{\text{DM}} \lambda }{v^3}\bm{v},
\end{equation}
where the subscript a means that it is due to the effect of accretion.
Combining Eqs. \eqref{f-aa}, \eqref{pk-pf}, \eqref{pk-ef} and \eqref{pk-wf}, we get
\begin{align}
    \label{pk-a-pf}
    \left\langle\frac{d p}{d \phi}\right\rangle_{\text{a}}&=-\frac{4\mu\rho_{\text{sp}} r_{\text{sp}}^{\alpha}\lambda}{M^2}p^{4-\alpha} g(e),\\
    \label{pk-a-ef}
    \left\langle\frac{d e}{d \phi}\right\rangle_{\text{a}}&=-\frac{4\mu\rho_{\text{sp}} r_{\text{sp}}^{\alpha}\lambda}{M^2}p^{3-\alpha} f(e),\\
    \label{pk-a-wf}
    \left\langle\frac{d \omega}{d \phi}\right\rangle_{\text{a}}&=0.
\end{align}

\begin{figure}[tbp]
    \centering
    \includegraphics[width=.47\textwidth,origin=c]{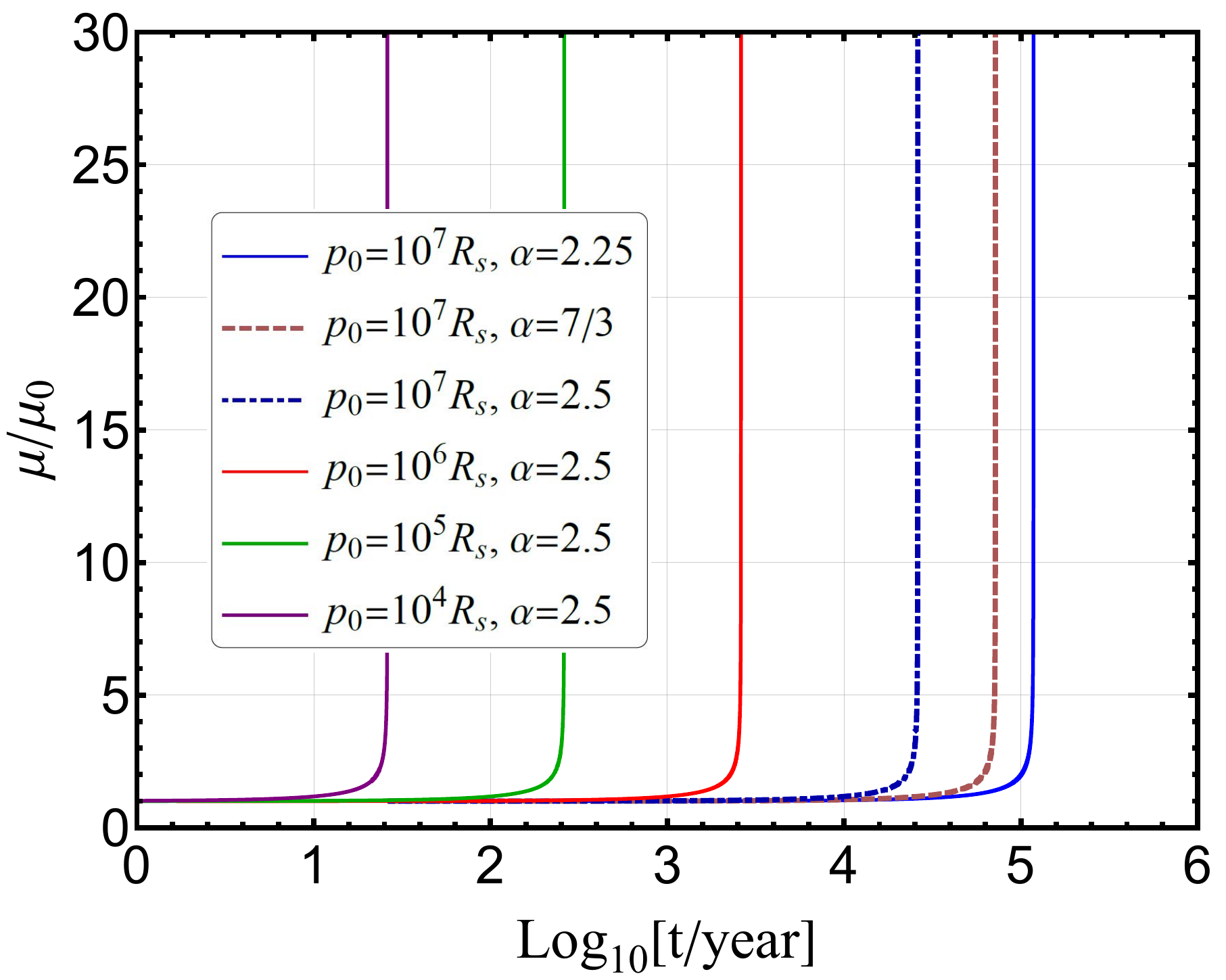}
    \caption{
    The growth of small BH' mass $\mu$  when only accretion is considered.
    The initial mass is $\mu_0=10M_{\odot}$.
    The initial eccentricity is $e_0=0.6$.
    Initial values of $p_0$ are $10^4 R_\text{s}$, $10^5 R_\text{s}$, $10^6 R_\text{s}$ and $10^7 R_\text{s}$.
    $\alpha$ is chosen as $2.25$, $7/3$, and $2.5$.
    }
    \label{fig:2.6}
\end{figure}

Equations. \eqref{pk-a-pf}, \eqref{pk-a-ef}, and \eqref{pk-a-wf} are the same as Eqs. \eqref{pk-df-pf}, \eqref{pk-df-ef} and \eqref{pk-df-wf} with $\lambda$ replacing $I_v$, so the effect of accretion is the same as DF. Solving the above orbital evolution equations, we get the growth of the small BH's mass as shown in Fig. \ref{fig:2.6}.
In Fig. \ref{fig:2.6}, the mass of the small BH grows from $p_0$ to $p=10 R_\text{s}$.
We can see that the mass of the small BH increases rapidly when $p\sim 10R_s$, it even reaches to thirty times that of the initial mass.
One reason for this is that the density of the DM minispike is larger as the small BH moves closer to the central IMBH.
Another reason is that there is a plenty of time for the small BH to grow when only the accretion is considered.
However, as we will see in the next section, when other factors such as the DF and the reaction of GWs are taken into account, there is not enough time for the small BH to become very large.

\subsection{Reaction of GWs}

The reaction of GWs on eccentric binaries was explored by Peters and Mathews \cite{Peters:1963ux,Peters:1964zz}.
The measurement of the orbital damping of pulsar binaries caused by GW reaction was then reported in \cite{Hulse:1974eb,Taylor:1979zz}.
The reaction of GWs can be calculated as a perturbation force \cite{Mino:1996nk,Pati:2000vt,Pati:2002ux} with the method of osculating orbit \cite{Pound:2007th,Pound:2010wa}.
In the harmonic gauge, the effect of reaction of GWs on the acceleration of the system can be written as \cite{Damour:1981bh,Poisson:2018gn}
\begin{align} \label{a-gw-rr}
\begin{split}
    \bm{a}_{\text{GW}}=& \frac{8}{5} \frac{G^2 M \mu}{c^5 r^3}\left[\left(3v^2+\frac{17}{3}\frac{G m}{r}\right)\dot{r}\bm{n}\right.\\
    &\left.-\left(v^2+3\frac{G m}{r}\right)\bm{v}\right].
\end{split}
\end{align}

Substituting Eq. \eqref{a-gw-rr} into Eqs. \eqref{pk-pf}, \eqref{pk-ef}, \eqref{pk-wf} and \eqref{pk-tf}, we obtain
\begin{align}
    \label{pk-rr-pf}
    \left\langle\frac{d p}{d \phi}\right\rangle_{\text{GW}}&=-\frac{8}{5}\eta\frac{(G m)^{5/2}}{c^5 p^{3/2}}\left(8+7e^2\right),\\
    \label{pk-rr-ef}
    \left\langle\frac{d e}{d \phi}\right\rangle_{\text{GW}}&=-\frac{8}{5}\eta\frac{(G m)^{5/2}}{c^5 p^{5/2}}\left(\frac{304}{24}e+\frac{121}{24}e^3\right),\\
    \label{pk-rr-wf}
    \left\langle\frac{d \omega}{d \phi}\right\rangle_{\text{GW}}&=0,\\
    \label{pk-rr-tf}
    \left\langle\frac{d t}{d \phi}\right\rangle_{\text{GW}}&= \sqrt{\frac{p^3}{G m}}(1-e^2)^{-3/2},
\end{align}
where $\eta=M\mu/(M+\mu)^2$,
and the subscript GW means that it is due to the effect of the reaction of GWs.
From Eqs. \eqref{pk-rr-pf}, \eqref{pk-rr-ef}, \eqref{pk-rr-wf} and \eqref{pk-rr-tf},
we see that the changes of $p$ and $e$ depend on $p$ as $p^{-3/2}$ and $p^{-5/2}$,
so the effect of the reaction of GWs is greater when the small BH moves closer to the central IMBH.
Unlike the DF, the reaction of GWs decreases both the orbital radius and the eccentricity.

\section{The Net Effect}
\label{Combined Effect}

In the previous section we discussed several perturbative forces and their effects on the orbital motion respectively.
In this section, we discuss the net effect of these perturbative forces.
Combining Eqs. \eqref{a-m+dm}, \eqref{f-df}, \eqref{ut-dm}, \eqref{eom-a-dm}, \eqref{a-gw-rr}, and \eqref{a-k}, we obtain
\begin{align}
    \label{a-cf1}
    \bm{a}_{\text{tot}}=& -\frac{G M_{\text{eff}}}{r^2}\bm{n}+\bm{a}_{\text{DM}}+\bm{a}_{\text{DF}}+\bm{a}_{\text{a}}+\bm{a}_{\text{GW}},\\
    \label{ut-dm1}
    \dot{\mu}\simeq & 4\pi G^2 \lambda\frac{{\mu}^2\rho_{\text{DM}}}{v^3},
\end{align}
where
\begin{align}
    \label{a-dm1}
    \bm{a}_{\text{DM}}=& -\frac{G F}{r^{\alpha-1}}\bm{n},\\
    \label{a-df1}
    \bm{a}_{\text{DF}}=& -\frac{4\pi G^2 \mu \rho_{\text{DM}} I_{v}}{v^3}\bm{v},\\
    \label{a-a1}
    \bm{a}_{\text{a}}\simeq & -\frac{4\pi G^2 \mu \rho_{\text{DM}} \lambda }{v^3}\bm{v},\\
    \label{a-rr1}
    \bm{a}_{\text{GW}}=& \frac{8}{5} \frac{G^2 M \mu}{c^5 r^3}\left[\left(3v^2+\frac{17}{3}\frac{G M}{r}\right)\dot{r}\bm{n}\right.\nonumber\\
    &\left.-\left(v^2+3\frac{G M}{r}\right)\bm{v}\right].
\end{align}

As discussed in the previous section,
the effect of these perturbation forces dominates at different orbital ranges.
For example, $\bm{a}_{\text{DM}}$ dominates at large orbital distance $p\gg 10^5 R_s$ only and is negligible at small orbital distance $p\ll 10^5 R_s$.
However, $\bm{a}_{\text{DF}}$, $\bm{a}_{\text{a}}$ and $\bm{a}_{\text{GW}}$ have more pronounced effects at small orbital distance  $p\ll 10^5 R_s$ than at large orbital distance  $p\gg 10^5 R_s$, and their effects on the orbit are accumulated.
Therefore, we consider the net effect at the large orbital distance $p\gg 10^5 R_\text{s}$ and at the small orbital distance $p\ll  10^5 R_\text{s}$ separately.

\subsection{Small orbital range $p\ll 10^5\ R_s$}
\label{near range}

In this subsection, we discuss the net effect at small orbital distance $p\ll 10^5\ R_s$.
As discussed above, the effect of the gravity of the DM minispike is negligible in this region, so
Eq. \eqref{a-cf1} becomes
\begin{equation}
    \label{a-cf2}
    \bar{\bm{a}} _{\text{tot}}\simeq -\frac{G M}{r^2}\bm{n}
    +\bm{a}_{\text{DF}}+\bm{a}_{\text{a}}+\bm{a}_{\text{GW}}.
\end{equation}
Substituting Eq. \eqref{a-cf2} into Eq. \eqref{pk-tf},
we obtain
\begin{align}\label{pk-ce-tf0}
    \begin{split}
        \left(\frac{d t}{d \phi}\right)_{\text{tot}}=&\sqrt{\frac{p^3}{G M}}\frac{1}{(1+e\cos\phi)^2}\\
        &\times\left\{1-\frac{8\pi \mu \rho_{\text{DM}}
        r_{\text{sp}}^{\alpha} p^{3-\alpha}\sin{\phi}(I_v+\lambda)M^{-2}}{e(1+2e\cos{\phi}+e^2)^{3/2}(1+e\cos{\phi})^{2-\alpha}}\right.\\
         &\left.-\frac{8\eta(G M)^{5/2}}{5c^5 p^{5/2}}\left[\left(\frac{3}{2}e^2+\frac{28}{3}+\frac{35}{6}e \cos\phi \right)\sin{2\phi}\right.\right.\\
         &\left.\left.+\frac{8}{e}\sin\phi+2e\sin\phi \right](1+e\cos\phi) \right\}.
    \end{split}
\end{align}
The second term in curly brackets of Eq. \eqref{pk-ce-tf0} is the correction from the DF and accretion,
and the third term is from the reaction of GWs.
Take the average of Eq. \eqref{pk-ce-tf0}, we get
\begin{equation}\label{pk-ce-tf}
    \left\langle\frac{d t}{d \phi}\right\rangle_{\text{tot}}=\sqrt{\frac{p^3}{G M}}(1-e^2)^{-3/2}.
\end{equation}
This form is the same as in Keplerian motion.

\begin{figure*}[htp]
    \centering
     \includegraphics[width=0.95\linewidth]{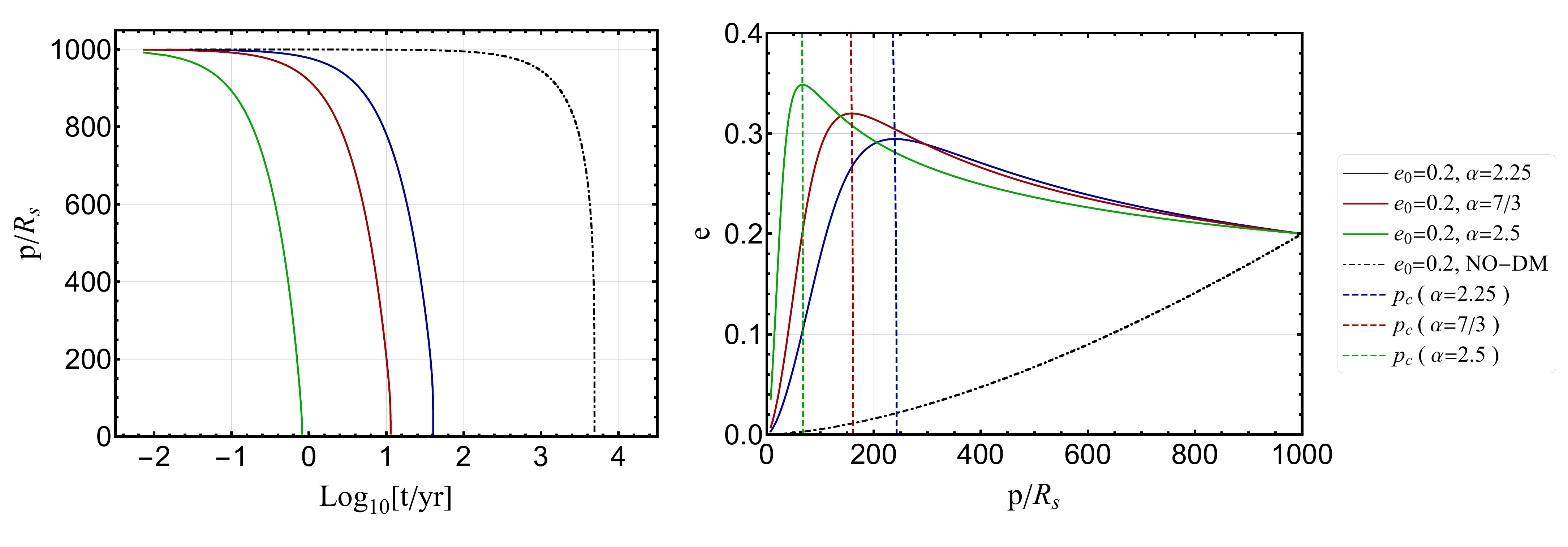}
     \includegraphics[width=0.95\linewidth]{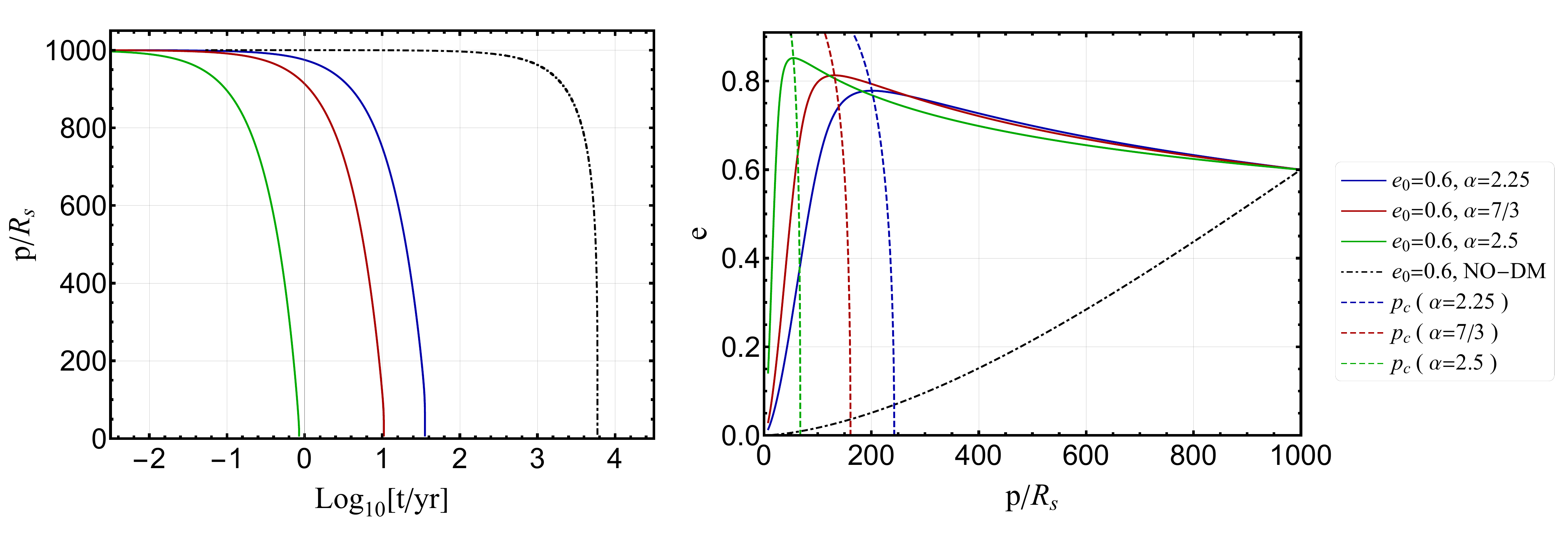}
     \includegraphics[width=0.95\linewidth]{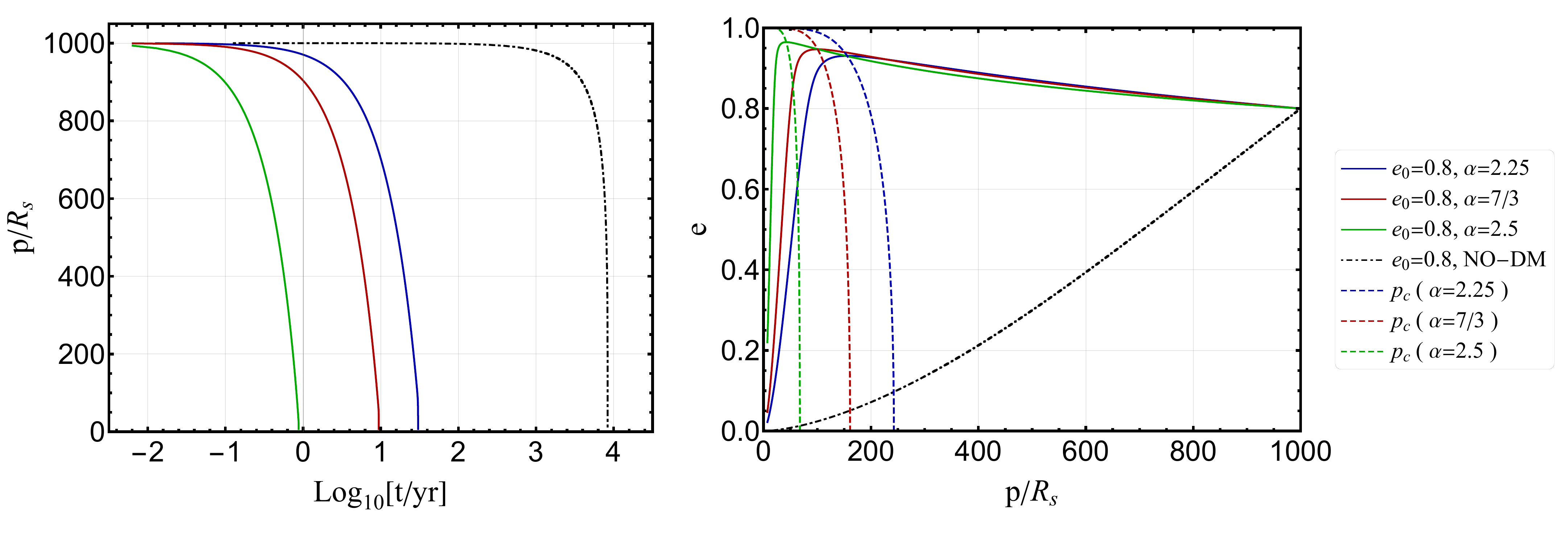}
    \caption{ The evolution of orbital parameters from the initial semi-latus rectum $p_0=10^3 R_\text{s}$ to $p=10 R_\text{s}$.
   The right panels show $e$ versus $p$ for different initial orbital parameters.
   The left panels show how $p$ evolves.
   We take the initial eccentricities as $0.2$, $0.6$, and $0.8$,
   and the values of $\alpha$ as $2.25$, $7/3$, and $2.5$.
   The black dashed lines are the cases with the same initial orbital condition but without DM.
   The color dashed lines in the right panels show how the critical radius $p_c$ change with $e$ for different values of $\alpha$.}
     \label{fig:3.01}
\end{figure*}

Substituting Eq. \eqref{a-cf2} into Eqs. \eqref{pk-pf}, \eqref{pk-ef}, and \eqref{pk-wf}, we obtain

    \begin{align}
        \label{pk-ce-pf}
        \left\langle\frac{d p}{d \phi}\right\rangle_{\text{tot}}=& -\frac{4\mu\rho_{\text{sp}} (r_{\text{sp}})^{\alpha}(I_v+\lambda)}{M^2}p^{4-\alpha} g(e)\\
        &-\frac{8}{5}\eta\frac{(G M)^{5/2}}{c^5 p^{3/2}}\left(8+7e^2\right),\nonumber\\
        \label{pk-ce-ef}
        \left\langle\frac{d e}{d \phi}\right\rangle_{\text{tot}}=& -\frac{4\mu\rho_{\text{sp}} (r_{\text{sp}})^{\alpha}(I_v+\lambda)}{M^2}p^{3-\alpha} f(e)\\
        &-\frac{8}{5}\eta\frac{(G M)^{5/2}}{c^5 p^{5/2}}\left(\frac{304}{24}e+\frac{121}{24}e^3\right),\nonumber\\
        \label{pk-ce-wf}
        \left\langle\frac{d \omega}{d \phi}\right\rangle_{\text{tot}}=& 0.
    \end{align}
From Eq. \eqref{pk-ce-wf},
we see that the net effect on the orbital precession is null.
In Eq. \eqref{pk-ce-pf}, $g(e)$ is always greater than zero,
so the orbital radius decreases with $\phi$.
However, the sign of the right-hand side of Eq. \eqref{pk-ce-ef} is uncertain,
so it is not clear whether the eccentricity increases or decreases with $\phi$.
Let the left-hand side of Eq. \eqref{pk-ce-ef} equal to zero,
we can define the critical radius
\begin{align} \label{pc-pe}
\begin{split}
    p_\text{c} &= \left[ \frac{-8(G M)^{5/2}\mu}{20 c^5 \rho_{\text{sp}}r_{\text{sp}}^{\alpha}(I_v +\lambda)f(e)}
    \left( \frac{304}{24}e+\frac{121}{24}e^3 \right)  \right]^{\frac{2}{11-2\alpha}}.
\end{split}
\end{align}
The value of $p_\text{c}$ depends on $e$ and $\alpha$.
When $p > p_\text{c}$, the effects of DF and accretion are stronger than the effect of the GW reaction, the eccentricity increases with $\phi$.
When $p < p_\text{c}$, the effect of GW reaction is more important, so the eccentricity decreases with $\phi$.
We show the evolutions of orbital parameters $p$ and $e$ for different initial values of $p_0$ and $e_0$ in Fig. \ref{fig:3.01}.
As shown in Fig. \ref{fig:3.01}, the presence of the DM minispike makes the orbit  decay more quickly.
The eccentricity increases slowly when $p > p_c$, and then decreases rapidly when $p\leq p_\text{c}$ due to the radiation of GWs.
Away from the central IMBH, the DF of DM minispike dominates over GW reaction, so $e$ increases with $\phi$.
If $\alpha$ is larger, the effect of DF becomes stronger,
it can increase the eccentricity up to smaller distance,
so the value of $p_c$ is smaller.

\begin{figure*}
   \includegraphics[width=.92\textwidth,origin=c]{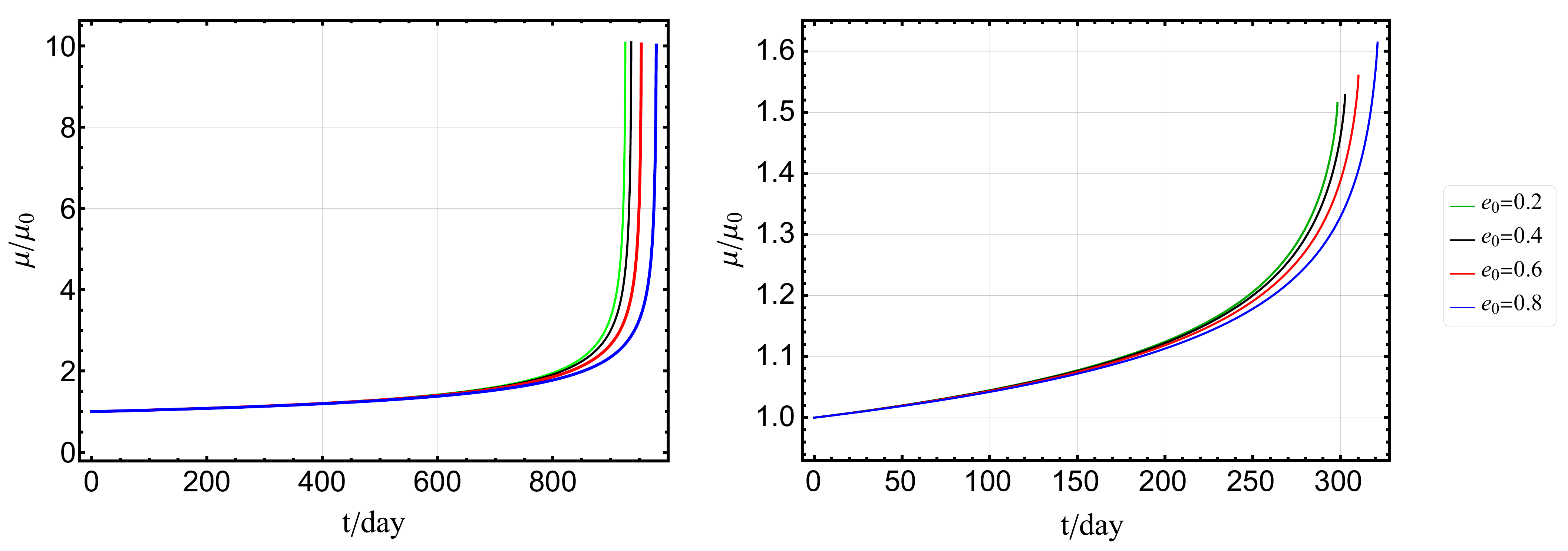}
   \caption{
   The growth of the small BH's mass from $p=10^3 R_\text{s}$ to $p=10 R_\text{s}$.
   The left panel shows the results with the effect of the accretion only.
   The right panel shows the results with the net effect.
   The initial eccentricities are chosen as $0.2$, $0.4$, $0.6$, and $0.8$,
   and the value of $\alpha$ is chosen as $2.5$.}
   \label{fig:3.01a}
\end{figure*}

From Eq. \eqref{ut-dm1}, we get
\begin{equation}
\label{ut-dm3}
\left\langle\dot{\mu}\right\rangle \simeq \frac{2 G^{1/2} {\mu}^2 \lambda\rho_{\text{sp}}}{M^{3/2}}
\frac{r_{\text{sp}}^{\alpha}}{p^{\alpha-3/2}}j(e),
\end{equation}
where 
\begin{equation}
    j(e)=\int_{0}^{2\pi} \frac{(1+e\cos{\phi})^{\alpha}}{(1+2e\cos\phi+e^2)^{3/2}}\,d\phi >0.\nonumber
\end{equation}
The growth of the small BH's mass from $p=10^3 R_s$ to $p=10 R_s$ is shown in Fig. \ref{fig:3.01a}.
The mass of the small BH could increase to $\sim 1.3-1.7$ times of the initial mass under the net effect,
as shown in the right panel of Fig. \ref{fig:3.01a}.
We see that if the initial value $e_0$ is larger,
it takes longer time for the IMRI to merge,
so the small BH accretes more DM and it becomes bigger.
The left panel of Fig. \ref{fig:3.01a} shows the change of the small BH's mass under the influence of the accretion only.
Comparing the results in Fig. \ref{fig:3.01a},
we see that the mass accretion by the net effect is much smaller than that by the effect of accretion only.
This is because the reaction of GWs becomes dominant when $p<100 R_s$ and the orbit decreases rapidly to merge,
so there is not enough time for the small BH to become big.

Combining Eqs. \eqref{pk-ce-pf}, \eqref{pk-ce-ef}, \eqref{ut-dm3}, and \eqref{pk-ce-tf0},
we can estimate the merging time of IMRIs.
Since the evolution time from $p=10 R_s$ to coalescence are a few hours or even less than one hour,
and the evolution from $p=10^3 R_s$ to $p=10 R_s$ would take many years,
so the evolution time of IMRIs from $p=10^3R_s$ to $p=10R_s$ can be approximated as the merger time from $p=10^3R_s$ to coalescence.
We show the evolution time of IMRIs from $p=10^3 R_s$ to $p=10 R_s$ with different initial eccentricities and different values of $\alpha$ in Table \ref{tab:3.1}.
Comparing with the results without DM,
the presence of DM minispike shortens the merger time greatly.
The larger the value of $\alpha$,
the faster the evolution of IMRIs with DM minispikes,
the shorter the time it takes to merge.
As the event rate of IMRIs/EMRIs is proportional to the inverse of the merger time \cite{Yue:2018vtk},
the existence of DM minispike greatly enhances the event rate of IMRIs.

\begin{table}
\begin{tabular}{ |p{0.8cm}|p{1.5cm}|p{1.5cm}|p{1.5cm}| p{1.5cm}| }
    \hline
    \multirow{1}{*}{$e$}   & No DM   & $\alpha=2.25$   & $\alpha=7/3$   & $\alpha=2.5$ \\
    \hline
   0     & 4829    & 41.0    & 11.5    & 0.813 \\
   0.2   & 4901    & 40.4    & 11.4    & 0.815 \\
   0.4   & 5178    & 38.6    & 11.1    & 0.826 \\
   0.6   & 5928    & 35.6    & 10.5    & 0.848 \\
   0.8   & 8354    & 30.3    & 9.5     & 0.879 \\
   0.9   & 12625   & 25.5    & 8.4     & 0.898 \\
    \hline
\end{tabular}
\caption{The time, in the unit of years, it takes the orbit evolving from $p=10^3 R_s$ to $p=10 R_s$.}
\label{tab:3.1}
\end{table}

In \cite{Yue:2017iwc}, the authors discussed the effect of DM minispike on the merger time for IMRIs in circular orbits (the case $e=0$) by considering the gravitational pull, DF, GW reaction, and accretion.
For IMRIs in eccentric orbits, only the effects of the DF and GW reaction on the merger time are considered in \cite{Yue:2019ozq}.
The effects of gravitational pull and accretion were not considered for eccentric IMRIs because they are difficult to calculate with that method used in \cite{Yue:2019ozq}.
With the osculating orbit method, here we consider the net effect of the gravitational pull, DF, GW reaction, and accretion for eccentric IRMIs.

Now we discuss the effect of perturbations on the GW waveform.
The quadrupole formula of GWs is
\begin{equation}
\label{h-jk}
h^{ij}=\frac{2G}{c^4 d_L} {\ddot{I}}^{ij},
\end{equation}
where $d_L$ is the luminosity distance to the GW source,
the dot denotes differential to the retarded time $\tau=t-d_L/c$,
and $I^{ij} $ is the mass quadrupole moment of the IMRI,
\begin{equation}
I^{ij}=\frac{M\mu}{M+\mu}r^i r^j.
\end{equation}
The plus and cross modes of GWs in the transverse-traceless gauge are
\begin{align}
    \label{h+}
    h_+=\frac{1}{2}(e^i_X e^j_X-e^i_Y e^j_Y)h_{ij},\\
    \label{hx}
    h_{\times} =\frac{1}{2}(e^i_X e^j_Y+e^i_Y e^j_X)h_{ij},
\end{align}
where $e_X$ and $e_Y$ are unit vectors perpendicular to the propagation direction $Z$ of GWs in the detector-adapted frame,
\begin{equation}
\label{h-jk-ma}
    \begin{split}
    h^{ij} \approx & \frac{4G M \mu}{c^4 R (M+\mu) }\left[ v^i v^j-\frac{G (M+\mu)}{r}n^i n^j\right.\\
    &\left.+\frac{4\pi G^2 \rho_{\text{DM}}}{v^3} \left(\frac{M\mu}{M+\mu}\right) \left(v^i r^j+ v^j r^i\right)\right],\\
    =& \frac{4 G^2 M \mu}{c^4 p R}\left[ -(1+e\cos\phi-e^2\sin^2\phi)n^i n^j \right.\\
    &\left.+\,e\,\sin\phi(1+e\cos\phi)(n^i k^j+k^i n^j)\right.\\
    &\left.+\,(1+e\cos\phi)^2 k^i k^j+A^{ij}\right],
    \end{split}
\end{equation}
$\bm{k}$ is the unit vector orthogonal to $\bm{n}$ and
\begin{align}\label{h-jk-mt}
    \begin{split}
    A^{ij}= & \frac{4\pi\rho_{\text{sp}}\,r_{\text{sp}}^{\alpha}M\mu}{(M+\mu)^3 p^{\alpha-3}}\frac{(1+e\cos\phi)^{\alpha}}{(1+2e\cos\phi+e^2)^{3/2}}\\
    &\times\left(\frac{2 e\sin\phi}{1+e\cos\phi}n^j n^k+k^i n^j +k^j n^i\right).
    \end{split}
\end{align}
The term $A^{ij}$ is the correction from the growth of the small BH and it is about $10^{-6}\sim 10^{-7}$ times smaller than the other terms.

\begin{figure*}[htbp]
    %\begin{tabular}{ccc}
   %\centering
   \includegraphics[width=.95\textwidth,origin=c]{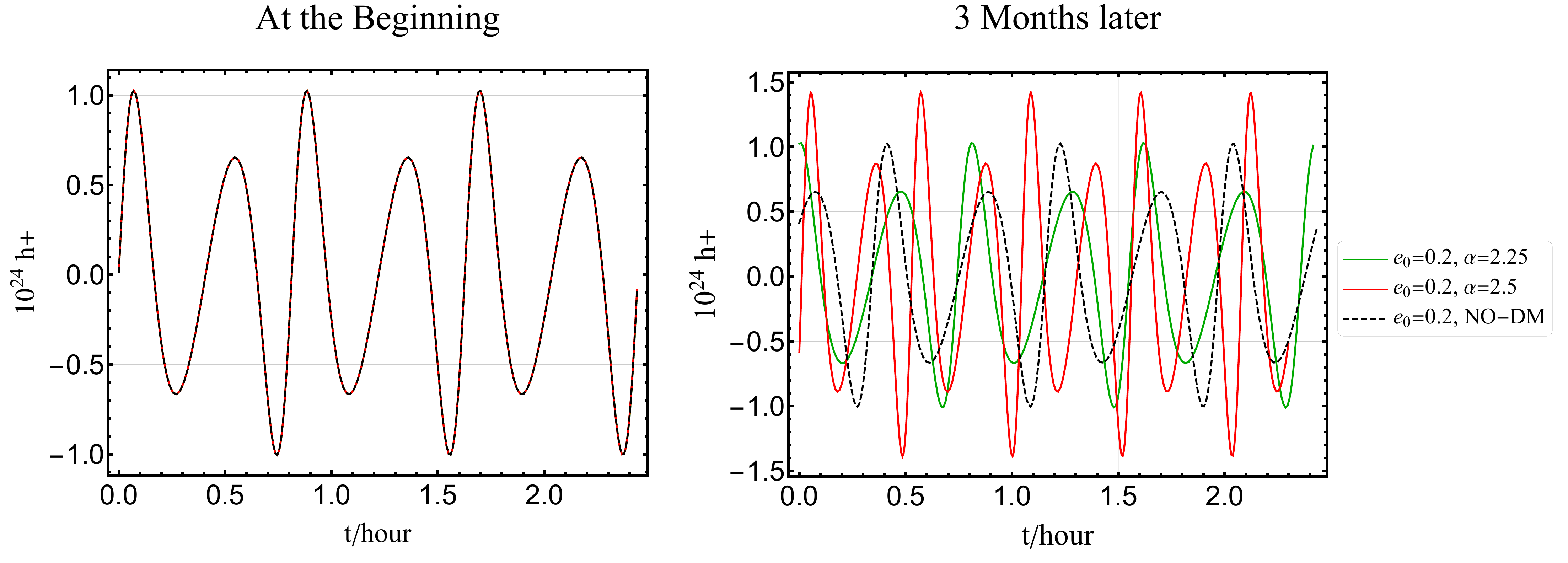}
   \includegraphics[width=.95\textwidth,origin=c]{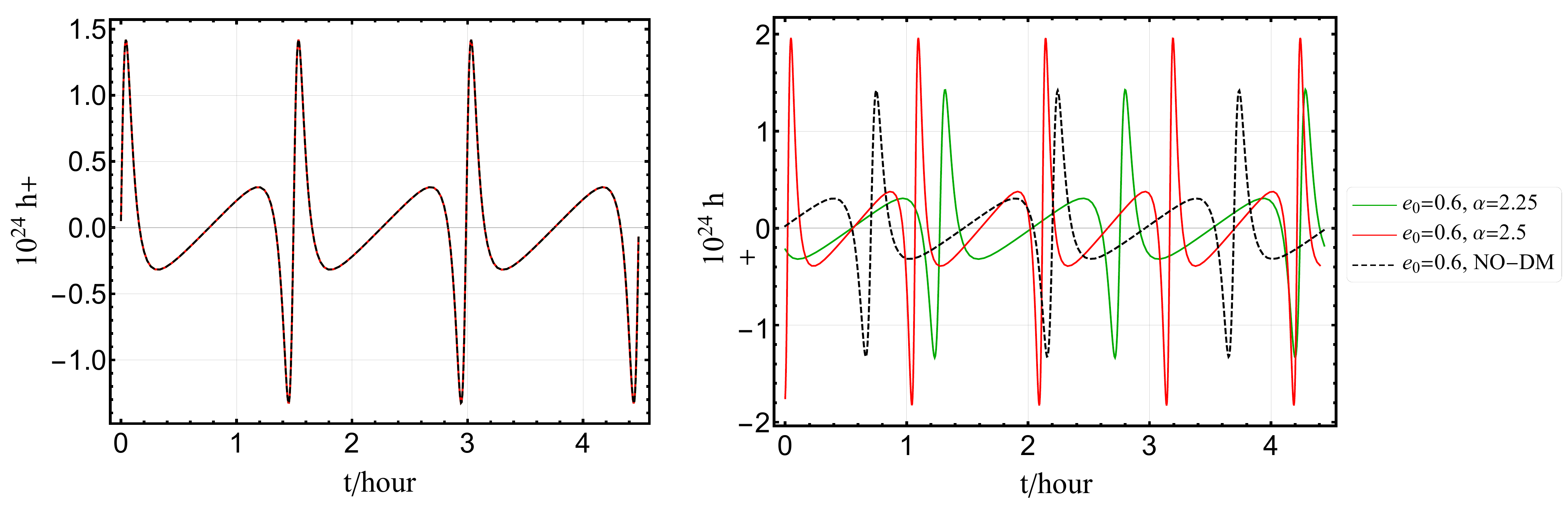}
   \includegraphics[width=.95\textwidth,origin=c]{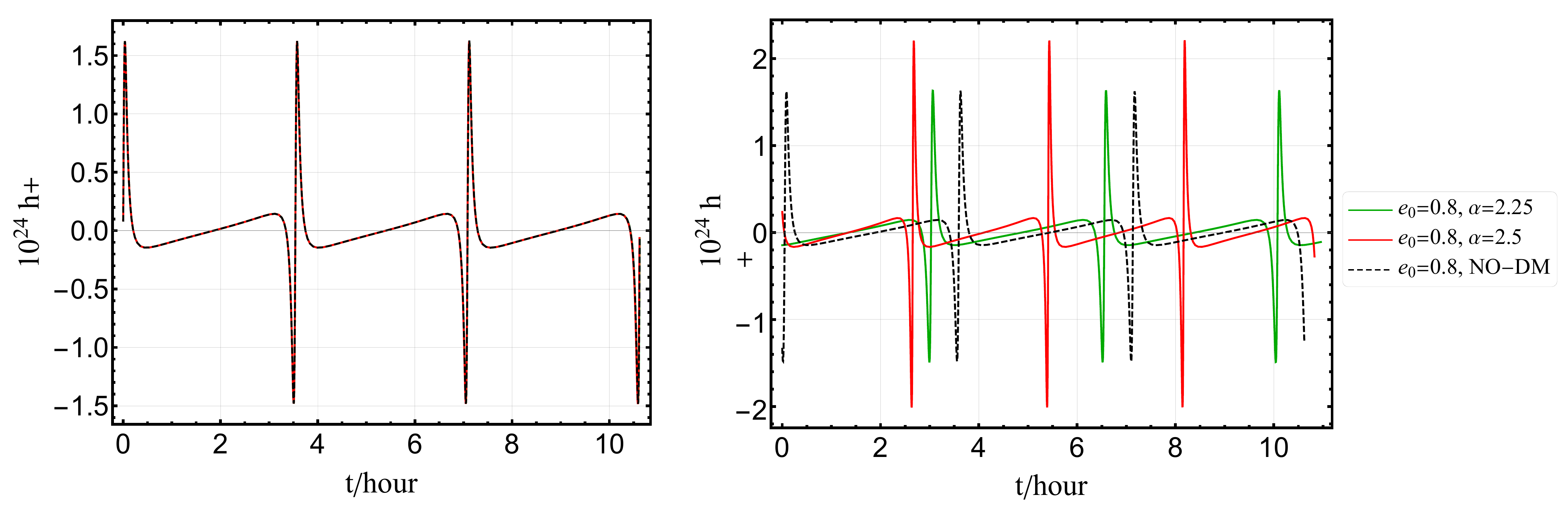}
   \caption{
   The time-domain plus mode GW waveform for IMRIs.
   The black-dashed lines are the waveforms without DM.
   The left panels show the initial waveforms.
   The right panels show the waveforms after three months.
   We take the initial eccentricity $e_0$ as $0.2$, $0.6$, and $0.8$ from the top to bottom panels, respectively,
   the initial semi-latus rectum as $p_0=10^3 R_s$, and the values of $\alpha$ as $2.25$ and $2.5$.
   The inclination angle $\iota=\pi/6$, the initial longitude of pericenter $\omega_0=0$, and $R=10^3\text{Mpc}$.
   }
    %\end{tabular}
    \label{fig:3.02}
\end{figure*}

The time-domain GW waveforms of the IMRI with different parameters are shown in Fig. \ref{fig:3.02}.
From Fig \ref{fig:3.02},
we see that initially the GW waveforms for IMRIs with and without DM minispike are the same.
Three months later, the GW waveforms are different.
The presence of DM minispike increases both the amplitude and frequency of GWs.
Therefore, long-time observation of GWs can be used to detect DM minispike and constrain the value of $\alpha$.

\begin{figure}[htbp]
   \includegraphics[width=.46\textwidth,origin=c]{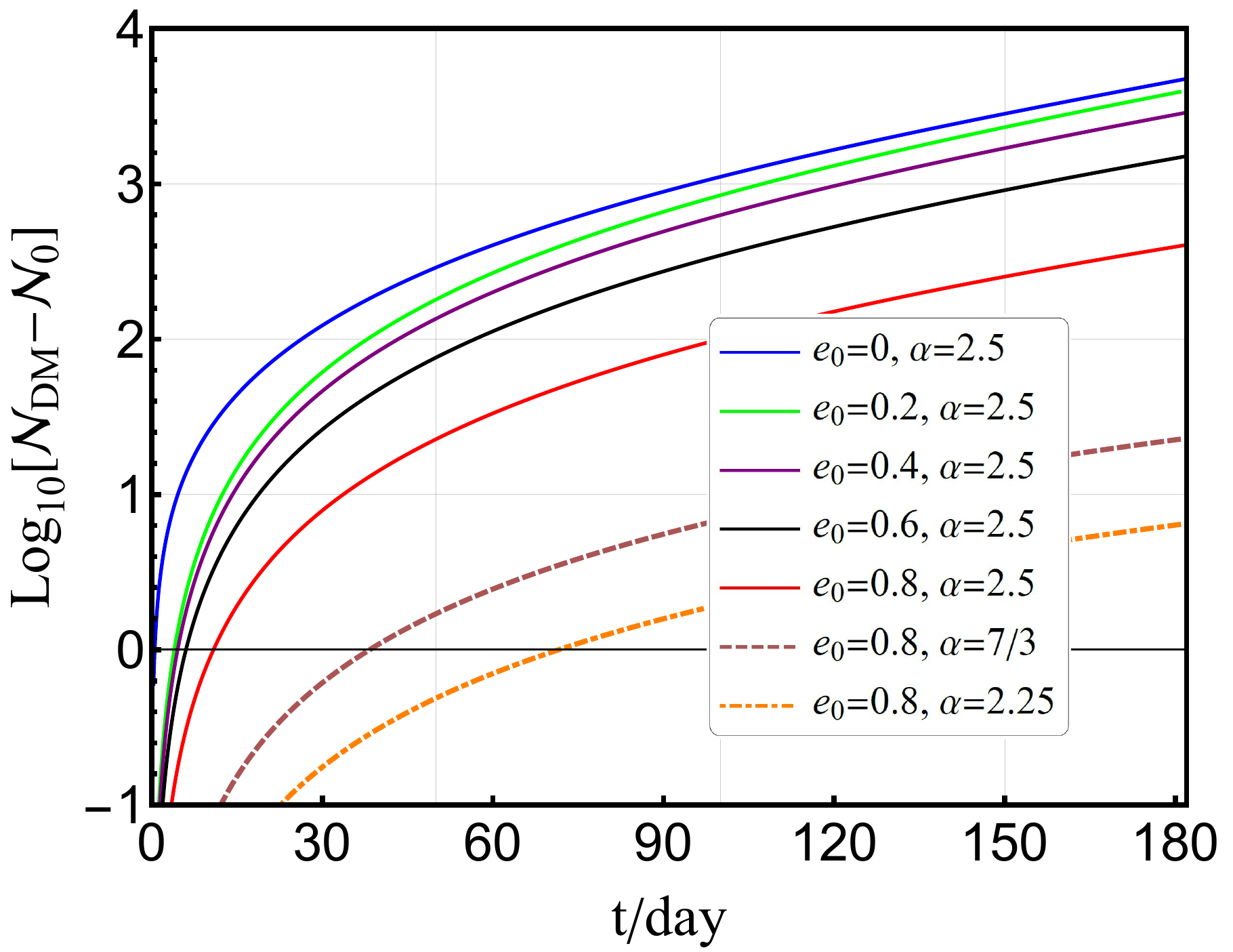}
   \caption{The difference $\mathcal{N}_\text{DM}-\mathcal{N}_0$ between the number of orbital cycles with and without DM minispikes
   accumulated during half-year evolution.
   We take the initial semi-latus rectum $p_0=10^3 R_s$.
   The initial eccentricities $e_0$ are chosen as $0$, $0.2$, $0.4$, $0.6$, and $0.8$ respectively, and the values of $\alpha$ are $2.25$, $7/3$, and $2.5$.}
    \label{fig:3.02a}
\end{figure}

To quantify the influence of DM minispike on GWs,
firstly we compare the number of orbital cycles accumulated for a long-time evolution of the IMRI as in \cite{Berti:2004bd,Kavanagh:2020cfn},
\begin{equation}\label{n-gw}
   \mathcal{N} =\frac{1}{2\pi}\int_{t_\text{i}}^{t_\text{f}} f(t)\,d\,t,
\end{equation}
where $f$ is the orbital frequency,
$t_\text{i}$ and $t_\text{f}$ are the initial and final time for the orbital evolution.
In Fig \ref{fig:3.02a}, we show the accumulated difference of the number of orbital cycles with and without DM minispike for the orbital evolution of six months.
We choose $f_\text{i}$ as the orbital frequency at $p_0=10^3 R_s$, 
and $f_\text{f}$ is the frequency after six months.
The difference between the number of orbital cycles with and without DM minispike is $\Delta\mathcal{N}(t)=\mathcal{N}_\text{DM}(t)-\mathcal{N}_\text{0}(t)$.
As shown in  Fig \ref{fig:3.02a}, 
we see there is significant difference in the number of cycles.
$\Delta\mathcal{N}(t)$ is always positive because the evolution of IMRIs with DM minispikes is faster than those without DM minispikes.
If $\alpha$ is larger,
the evolution of the IMRIs with DM minispike is faster,
so $\Delta\mathcal{N}(t)$ becomes larger.
When $\alpha=2.5$, the difference between the number of orbital cycles can be much more than $10^3$.

Then we calculate the SNR with LISA for GWs emitted from IMRIs with and without DM minispike and the mismatch between these two GWs.
Given two signals $h_1(t)$ and $h_2(t)$, 
we define the inner product $(h_1|h_2)$ as
\begin{equation}\label{overlap}
    (h_1|h_2)=2\int_0^{+\infty } \frac{\tilde{h}_1(f)\tilde{h}_2^*(f)+ \tilde{h}_2(f)\tilde{h}_1^*(f)}{ S_h(f)} \, df, 
\end{equation}
where $\tilde{h}(f)$ is the Fourier transformation of the time series $h(t)$, $h^*$ denotes the complex conjugation
and $S_h$ is the one-sided noise power spectral density (PSD).
The SNR for a signal $h$ is $\sqrt{(h|h)}$. 
The PSD of LISA is \cite{Robson:2018ifk}
\begin{align}\label{psd-lisa}
    \begin{split}
    S_h(f) =& \frac{S_x}{L^2}+\frac{2S_a \left[1+\cos^2(2\pi\,f L/c)\right]}{(2\,\pi f)^4 L^2}\\
    &\times\,\left[1+\left(\frac{4\times 10^{-4}\text{Hz}}{f}\right) \right],
    \end{split}
\end{align}
where $\sqrt{S_a}=3\times 10^{-15}\ \text{m s}^{-2}/\text{Hz}^{1/2}$ is the acceleration noise,
$\sqrt{S_x}=1.5\times 10^{-11}\ \text{m/Hz}^{1/2}$ is the displacement noise and $L=2.5 \times 10^6\text{ km}$ is the arm length of LISA \cite{Audley:2017drz}. 
The overlap between two GW signals is quantified as \cite{Babak:2006uv}
\begin{equation}
    \mathcal{O}(\tilde{h}_1,\tilde{h}_2)= \frac{(\tilde{h}_1|\tilde{h}_2)}{\sqrt{(\tilde{h}_1|\tilde{h}_1)(\tilde{h}_2|\tilde{h}_2)}},
\end{equation}
and the mismatch between two signals is defined as
\begin{equation}\label{mismatch}
    \text{Mismatch}=1-\mathcal{O}_\text{max}(\tilde{h}_1,\tilde{h}_2),
\end{equation} 
where the maximum is evaluated with respect to the time shift and the orbital-phase shift.
The mismatch is zero if two signals are identical.
Two signals are considered experimentally distinguishable if their mismatch is larger than $d/(2\rho^2)$, where $d$ is the number of intrinsic parameters of the GW source \cite{Flanagan:1997kp,Lindblom:2008cm,Buonanno:2002ft}.

Choosing different initial eccentricity $e_0$ at $p_0=10^3 R_s$ and taking $\alpha=7/3$,
we calculate the SNR for and the mismatch between GWs from eccentric IMRIs with and without DM minispike at the luminosity distance $d_L=100$ Mpc with one year integration time prior to $p=10R_s$,
we also calculate the maximum detectable distance $D_\text{max}$ by fixing the SNR to be $12$ \cite{Amaro-Seoane:2007osp},
these results are summarized in Table \ref{t-s1}.
From Table \ref{t-s1}, we see that without DM minispikes, 
the values of SNR are almost the same for eccentric IMRIs with different $e_0$. 
But the values of SNR are different for eccentric IMRIs with different $e_0$ when DM minispikes are present. 
So the values of $e_0$ affect the SNR of IMRIs with DM minispikes.
Note that the SNR and the maximum detectable distance increase as $e_0$ becomes larger initially, but then they decrease if $e_0$ is too big.   
The mismatch between GWs from eccentric IMRIs with and without DM minispike is much larger than $d/(2\,\text{SNR}_{0}^2)=4.3\times10^{-3}$ for all cases.
Thus we can detect DM minispike with LISA.
The maximum detectable distance with LISA can be estimated as $D_\text{max}=\text{SNR}_D/12\times 100\text{ Mpc}$, which is $\sim 300$ Mpc.

\begin{table}
    \begin{tabular}{ |p{0.8cm}|p{1cm}|p{1cm}|p{1.6cm}|p{1.0cm}| }
        \hline
        $e_0$ & $\text{SNR}_0$  & $\text{SNR}_D$  & Mismatch & $D_\text{max}$  \\
        \hline
        0.2  & 34.13  & 42.97  & 0.99992  & 358.1 \\
        \hline
        0.4  & 34.19  & 47.74  & 0.99944  & 397.8 \\
        \hline
        0.6  & 34.44  & 36.71  & 0.99965  & 305.9 \\
        \hline
    \end{tabular}
    \caption{
    The results of the SNR for and the mismatch between GWs from eccentric IMRIs with and without DM minispike for different initial values of the eccentricity at $p_0=10^3 R_s$. The luminosity distance is chosen as $d_L=100$ Mpc.
    We take one year integration time prior to $p=10 R_s$.
    $\text{SNR}_0$ and $\text{SNR}_D$ are the SNR for GWs from eccentric IMRIs without and with DM minispike, respectively.
    $D_\text{max}$, in units of Mpc, is the maximum detectable distance with SNR$=12$.
    }
    \label{t-s1}
\end{table}

To assess the detector's ability to constrain the parameter $\alpha$, we can perform parameter estimation for $\alpha$ using the FIM method \cite{Finn:1992wt,Cutler:1994ys,Poisson:1995ef}.
Unfortunately, there is no analytical waveform for eccentric IMRIs with DM minispike. 
For quasi-circular orbits, we can derive analytical waveforms \cite{Eda:2014kra,Yue:2017iwc} and the details are presented in the Appendix \ref{APPENDIX2}. 
With the analytical waveform, we estimate the parameter errors with the FIM method and the result is shown in Table \ref{tab:3.3}.

\begin{table}
    \begin{tabular}{ |p{1.1cm}|p{0.8cm}|p{0.8cm}|p{1.8cm}|p{1.7cm}|p{1.4cm}| }
        \hline
        \multirow{1}{*}{$\alpha$} & $\Delta\Phi_c$  & $\Delta t_c$  & $\Delta\ln\mathcal{M}_c(\%)$  & $\Delta\alpha$  & $\Delta\ln\kappa(\%)$  \\
        \hline
      No DM  & 0.496  & 2.08  & $4.27\times 10^{-7}$  & -                     & -        \\
      2.25   & 4.40   & 5.25  & $1.96\times 10^{-4}$  & $1.16\times 10^{-6}$  & 0.000303 \\
      7/3    & 1.88   & 1.73  & $9.91\times 10^{-4}$  & $1.64\times 10^{-6}$  & 0.000414 \\
      2.5    & 4.88   & 7.46  & $1.97\times 10^{-1}$  & $6.34\times 10^{-5}$  & 0.0207   \\
        \hline
    \end{tabular}
\caption{The estimated errors of the parameters with LISA for IMRIs with DM minispike in circular orbits.
We choose four-year observation time before the coalescence and SNR$=10$ \cite{Will:1994fb,Yagi:2009zm}.
The parameter $\kappa$ which is related to DM parameters $\rho_\text{sp}$ and $r_\text{sp}$ is defined in Eq. \eqref{kappa}.}
    \label{tab:3.3}
\end{table}

From Table \ref{tab:3.3}, we see that the error of $\alpha$ is in the order of $10^{-5}\sim 10^{-6}$.
If we consider eccentric orbits, 
we expect that the error $\Delta\alpha$ will be larger due to the addition of the eccentricity parameter $e$, but it should still be small.
Therefore, it is possible to detect DM minispike with LISA, Taiji, and Tianqin,
and place stringent constraint on the DM parameter $\alpha$.
The constraint on $\alpha$ can help us to understand the type of DM \cite{Hannuksela:2019vip}.

\subsection{Large orbital range $p\gg 10^5 R_s$}
\label{far range}

At large orbital distance, the small BH can also be compact object.
As discussed in Sec. \ref{DM-influence},
the orbital precession caused by the gravity of the DM minispike is much greater than that caused by the higher-order effect of gravitational interaction in the far region $p\gg 10^5 R_s$.
We also find that at a large orbital distance $p\gg 10^5 R_s$,
the effect of the DM minispike's gravity is much greater than those of the DF,
the accretion and the reaction of GWs.
Thus at large orbital distance, 
we mainly consider the effect of DM minispike's gravity.

\begin{figure*}[htbp]
   \includegraphics[width=.95\textwidth]{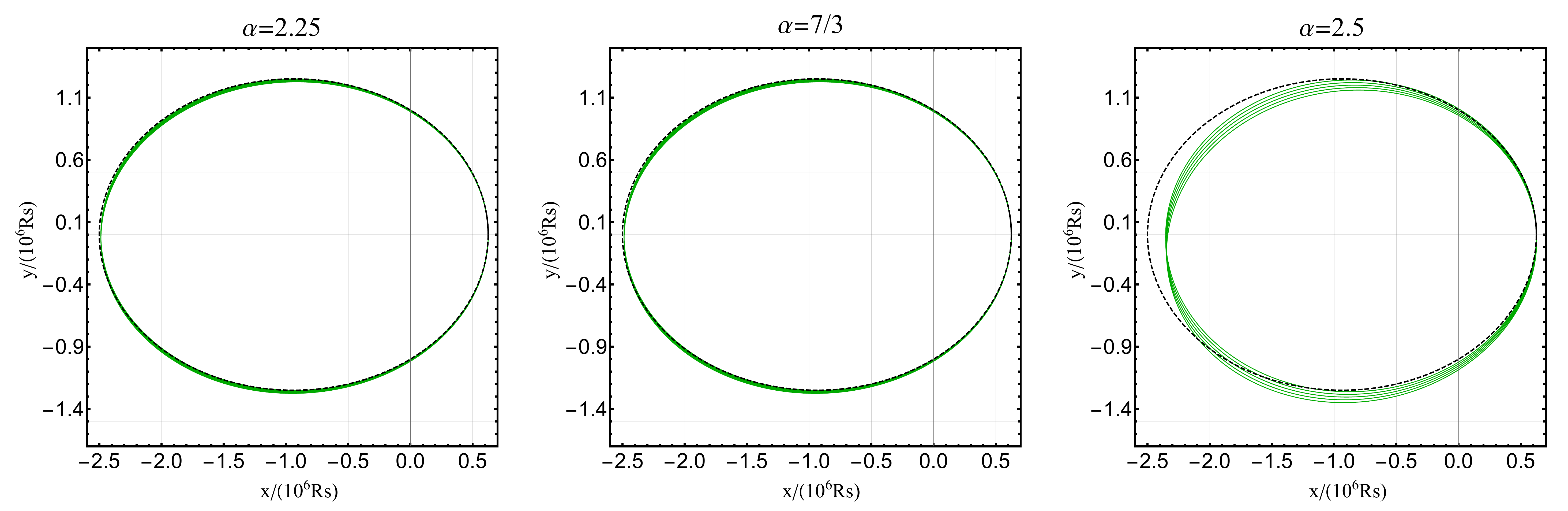}
   \includegraphics[width=.95\textwidth]{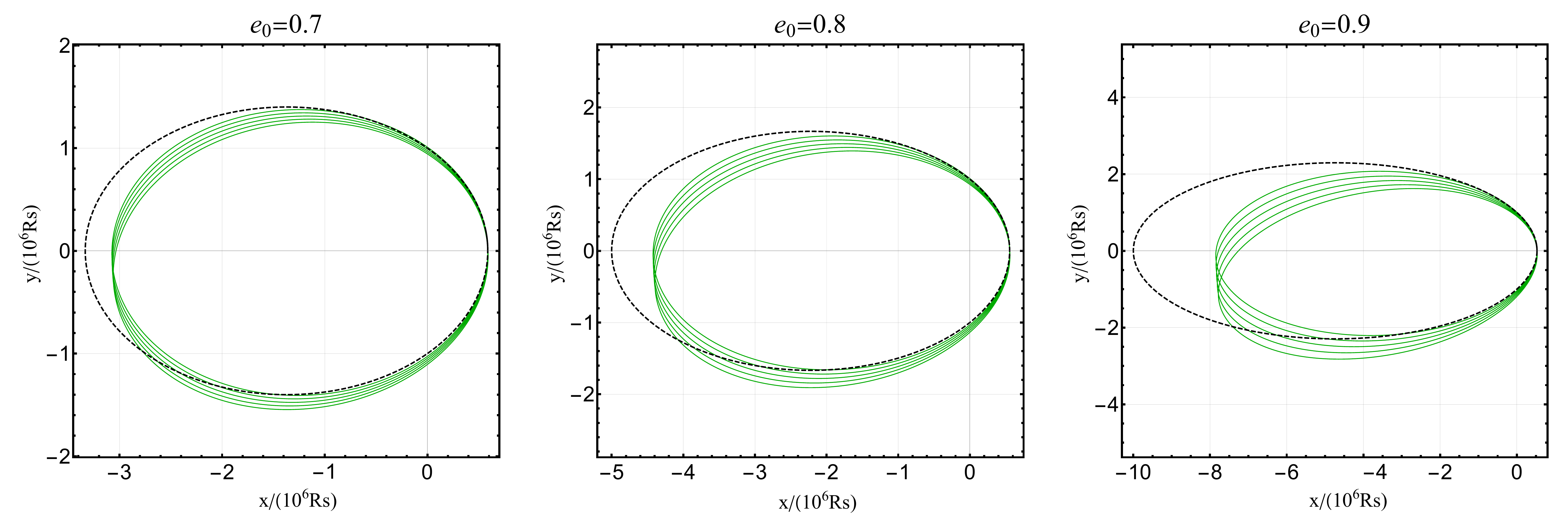}
   \caption{
   The orbital motion of IMRIs in large orbital distance.
   In the top panels, the initial eccentricity $e_0$ is $0.6$, the initial semi-latus rectum $p_0$ is $10^6R_s$, and the values of $\alpha$ are $2.25$, $7/3$, and $2.5$ from the left to the right panels.
   In the bottom panels, we take $p_0=10^6R_s$ and $\alpha=2.5$.
   The initial values of $e_0$ are $0.6$, $0.8$, and $0.9$ from the left to the right panels.
   The solid green lines are for IMRIs with DM and the dashed black lines are for IMRIs without DM.
   }
   \label{fig:3.03}
\end{figure*}

Combining Eqs. \eqref{dm-pf}, \eqref{dm-ef}, \eqref{dm-wf}, and \eqref{dm-tf}, 
we get the information about the IMRI's motion.
We plot the orbital motion of the binary with different parameters in Fig.~\ref{fig:3.03}.
As shown in the top panel in Fig.~\ref{fig:3.03},
the existence of DM minispike leads to the orbital precession.
If $\alpha$ is larger, i.e., the DM minispike is denser,
then the orbital precession is bigger.
In the bottom panel,
we see that larger eccentricity also causes bigger orbital precession.

\begin{figure*}[htbp]
\includegraphics[width=.96\textwidth,origin=c]{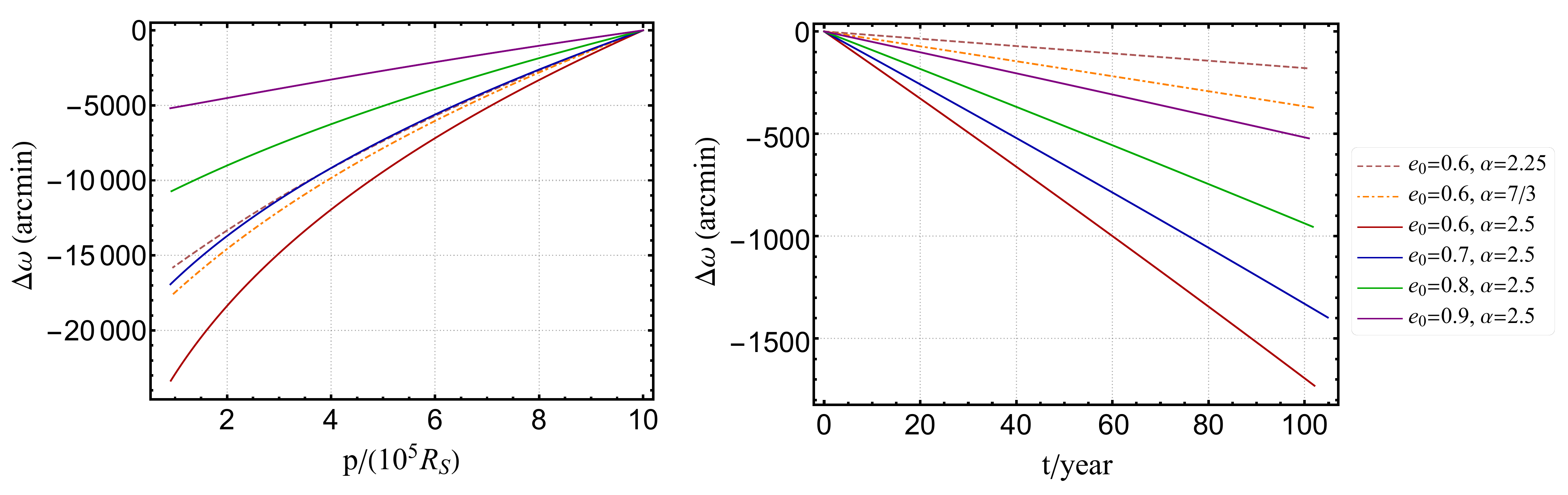}
  \caption{The total orbital precession $\Delta\omega_\text{tot}$ under the net effect.
   The left panel shows the orbital precession accumulated from $p=10^6 R_s$ to $p=10^5 R_s$.
   The right panel shows the orbital precession accumulated over 100 years starting with $p_0=10^6 R_s$.
   The initial eccentricities are chosen as $0.6$, $0.7$, $0.8$, and $0.9$,
   and the values of $\alpha$ are chosen as $2.25$, $7/3$, and $2.5$.}
   \label{fig:3.03a}
\end{figure*}

Since $W_\text{DM}<0$ and $W_\text{rp}>0$,
the orbital precession $\Delta\omega_\text{DM}$ induced by the gravity of DM minispike is negative,
and the orbital precession induced by the high-order effect of the gravity $\Delta\omega_\text{rp}$ is positive,
so the sign of total orbital precession $\Delta\omega_\text{tot}=\Delta\omega_\text{DM}+\Delta\omega_\text{rp}$ is uncertain.
In Fig. \ref{fig:3.03a}, we show the total orbital precession for different eccentricities and different values of $\alpha$ under the net effect of the DM minispike's gravity, the DF, the accretion and the reaction of GWs.
As shown in Fig. \ref{fig:3.03a},
we see that $\Delta\omega_\text{tot}<0$ because the effect of DM minispike's gravity is greater than the high-order effect of the gravity when $p>10^5 R_s$.
When the value of $\alpha$ and the initial eccentricity are larger, 
the IMRI evolves more quickly,
the time it takes for IMRIs evolving from $p=10^6 R_s$ to $p=10^5 R_s$ is smaller.
Thus the precession accumulated over the same time is larger if the value of $\alpha$ and the initial eccentricity are bigger.
Therefore, observations of orbital precession may disclose the DM minispike and its profile.

At large orbital distance $ 10^5 R_s-10^7 R_s$,
the frequency of GWs emitted by the IMRI is in the range $ 10^{-6}\text{ Hz}-10^{-9}\text{ Hz}$ which is the sensitivity band of pulsar timing array (PTA) \cite{foster1990constructing,NANOGrav:2020bcs}. 
At a large orbital distance, the plus and cross modes are
\begin{align}
    \label{dm-gw-x}
    h_{+}=& -\frac{2G^2 M\mu}{c^4\,p\,R} \left(1+\cos^2{\iota }\right)\bigg\{ \bigg[\cos\left(2\phi+2\omega\right)\nonumber\\
    &+\frac{5e}{4}\cos\left(\phi+2\omega\right)+\frac{e}{4}\cos\left(3\phi+2\omega\right)-\frac{e^2}{2}\cos{2\omega} \bigg]\nonumber\\
    &+\frac{e}{2}\sin^2{\iota }\left(\cos{\phi}+e\right)\bigg\},\\
    \label{dm-gw-+}
    h_{\times }=& -\frac{4G^2 M\mu}{c^4\,p\,R}\cos{\iota }\left[\sin\left(2\phi+2\omega\right)+\frac{5e}{4}\sin\left(\phi+2\omega\right)\right.\nonumber\\
    &\left.+\frac{e}{4}\sin\left(3\phi+2\omega\right)+\frac{e^2}{2}\sin{2\omega} \right].
\end{align}

\begin{figure*}[htbp]
    %\begin{tabular}{ccc}
   %\centering
   \includegraphics[width=.98\textwidth]{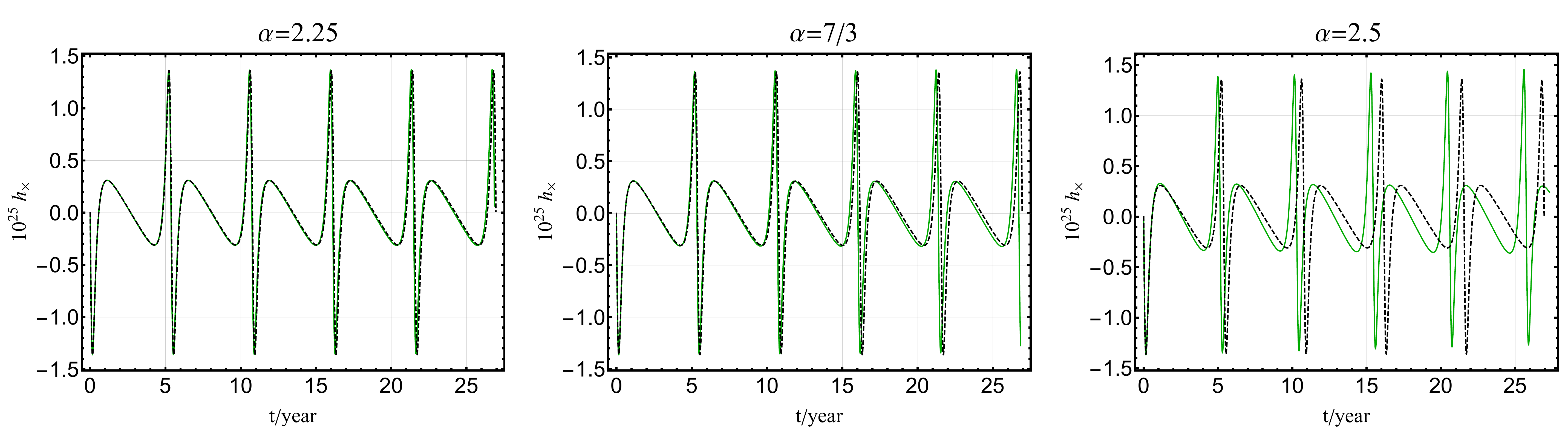}
   \includegraphics[width=.98\textwidth]{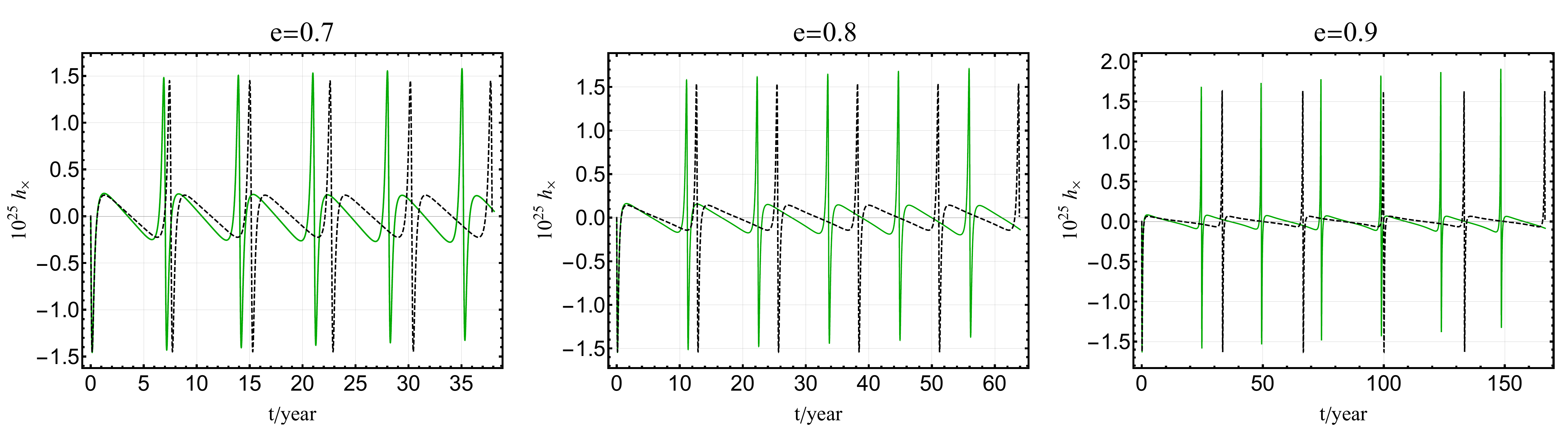}
   \caption{
   The time-domain plus mode waveforms for IMRIs in large orbital distances.
   We take the inclination angle $\iota=\pi/6$, the initial longitude of pericenter $\omega_0=\pi/4$ and the luminosity distance to the source $d_L=10\text{Mpc}$.
   In the top panel, the initial eccentricity $e_0$ is $0.6$, the initial semi-latus rectum $p_0$ is $10^6 R_s$, and the values of $\alpha$ are $2.25$, $7/3$, and $2.5$ from the left to the right panels.
   In the bottom panel, we take $p_0=10^6 R_s$ and $\alpha=2.5$, and the values of $e_0$ are $0.2$, $0.6$, and $0.8$ from the left to the right panels.
   The solid green lines are for IMRIs with DM and the dashed black lines are for IMRIs without DM.
   }
    %\end{tabular}
    \label{fig:3.04}
\end{figure*}

Using the orbital motion, we get the time-domain GW waveforms as shown in Fig. \ref{fig:3.04}.
There are obvious phase difference between the waveforms with and without DM minispike.
As discussed above, 
larger $\alpha$ and $e$ cause bigger orbital precession
and shorter orbital period,
so we see larger phase shift of GWs in Fig. \ref{fig:3.04}.
Although the amplitude of GWs from IMRIs in large orbital distance is small, 
these GWs may be observed by PTA and provide constraint on the profile of the DM minispike in the future \cite{Lee:2011et,Wang:2020hfh,Hobbs:2014tqa,Janssen:2014dka,Bailes:2021tot}.

\section{conclusions }
\label{conclusions}

The existence of the DM minispike affects the orbital motion of the IMRI consisting of the IMBH and a stellar mass BH or other compact object.
The orbital motion of the IMRI is affected by several factors, such as
the gravity of both the IMBH and the DM minispike, the DF from the DM minispike, 
and the accretion of the small BH, and the gravitational radiation reaction.
We find that the gravity of the DM minispike causes orbital precession,
but its effect is negligible at small orbital distances, $p\ll 10^5 R_s$.
At large orbital distances, $p\gg 10^5 R_s$,
the main contribution to the orbital precession comes from the gravity of the DM minispike.
The DF and accretion of the small BH decrease the orbital radius and increase the eccentricity.
However, the reaction of GWs decreases both the radius and the eccentricity.
At large orbital distances, $p\gg 10^5 R_s$, the orbital precession induced by the gravity of the DM minispike is large and it can be as large as 1500 arcmin over 100 years.

The effects of the DF, the accretion, and the reaction of GWs are important at small orbital distances, $p\ll 10^5 R_s$.
At small orbital distances, $p\ll 10^5 R_s$,
the net effects of the DF, the accretion and the reaction of GWs
are that the orbit decays faster,
the eccentricity increases for a long time then decreases rapidly,
the mass of the small BH increases to $\sim 1.3-1.7$ times,
the merger time is shortened greatly,
and the amplitude and frequency of GWs emitted become larger.
The GW waveforms from IMRIs with and without a DM minispike have a significant phase difference, which leads to a significant difference in the number of GW cycles accumulated over long-time evolution. The accumulated difference between the number of orbital cycles can reach $10^4$ over the half of a year.
Without DM minispike, the SNR is almost the same for eccentric IMRIs with different $e_0$, but the SNR is different for eccentric IMRIs with different $e_0$
when DM minispike is present.
The SNR and the maximum detectable distance increase as $e_0$ becomes larger initially, but then they decrease if $e_0$ is too big.
Therefore, the value of $e_0$ affects the SNR and the maximum detectable distance for eccentric IMRIs with DM minispikes.
The mismatch between GWs from eccentric IMRIs with and without a DM minispike is 
almost 1 which is much larger than $d/(2\,\text{SNR}_{0}^2)$.
The FIM analysis for IMRIs with DM minispikes in circular orbits shows that 
it is possible to detect DM minispikes with LISA, Taiji, and Tianqin,
and place stringent constraint on the DM parameter $\alpha$.

In conclusion, the observations of orbital precession and GWs may disclose the DM minispike and its density profile.

\begin{acknowledgments}

The computation is completed in the HPC Platform of Huazhong University of Science and Technology.
This research is supported in part by the National Key Research and Development Program of China under Grant No. 2020YFC2201504, the National Natural Science
Foundation of China under Grant No. 11875136 and No. 12147120, and the Major Program of the National Natural Science Foundation of China under Grant No. 11690021.
D. L. gratefully acknowledges the financial support from China Postdoctoral Science Foundation under Grant No. 2021TQ0018.

\end{acknowledgments}

\appendix
\section{The method of osculating orbital perturbation}
\label{APPENDIX}

In this appendix we introduce the method of osculating orbit, which was initially devised by Euler and Lagrange to treat Keplerian perturbation problems such as the three-body problem and external forces of the system.
In the method, the motion is always described by a sequence of Keplerian orbit with the orbital constants evolving under the perturbation \cite{Poisson:2018gn}.

\begin{figure}[H]
    \centering 
    \includegraphics[width=.46\textwidth]{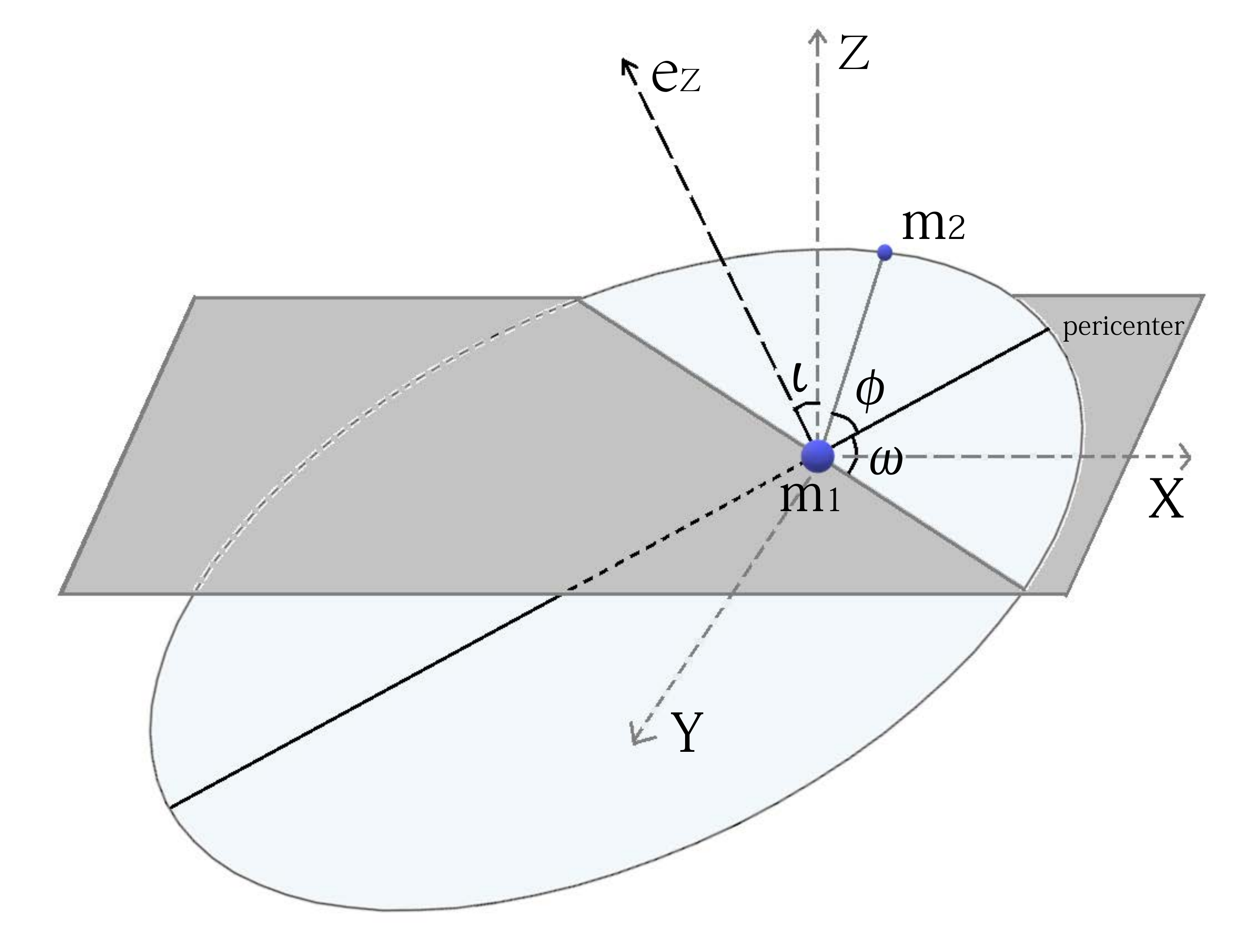}
    \caption{Orbital motion viewed in the fundamental reference frame.}
    \label{fig:x.1}
\end{figure}

We illustrate the Keplerian orbit in Fig. \ref{fig:x.1}.
In the fundamental frame with coordinates $(X, Y, Z)$, we adopt that the $Z$-axis points from the GW detector to the GW source.
In Fig. \ref{fig:x.1}, $\iota$ is inclination angle between the $X-Y$ plane and the orbital plane.
$\omega$ is the longitude of the pericenter which is the angle between the intersecting line of the two planes and the direction to the pericenter, 
$\phi$ is the angle between the separation vector $\bm{r}$ and the direction to the pericenter.

The relative acceleration of two bodies in a Keplerian orbit is
\begin{equation} \label{a-k}
    \bm{a}=-\frac{Gm}{r^2}\bm{n} +\bm{f},
\end{equation}
where $\bm{f}$ is a perturbing force per unit mass,
\begin{equation} \label{p-f}
    \bm{f} =\mathcal{R}\,\bm{n} +\mathcal{S}\,\bm{k}+\mathcal{W}\,\bm{e}_z,
\end{equation}
$\bm{n}=\bm{r}/r$ is the unit vector along the radius, $\bm{k}$ is the unit vector orthogonal to $\bm{n}$ and $\bm{e}_z$ is the normal vector of the orbital plane.

The equations of the osculating orbital elements are 
\begin{align}
    \label{pk-pt}
    \frac{d p}{d t}=& 2\sqrt{\frac{p^3}{Gm}}
    \frac{1}{1+e\cos\phi}\mathcal{S},\\
    \label{pk-et}
    \frac{d e}{d t}=& \sqrt{\frac{p}{Gm}}
    \left[\sin\phi\,\mathcal{R}+\frac{2\cos\phi+e(1+\cos^2 \phi)}{1+e\cos\phi}\mathcal{S}\right],\\
    \label{pk-wt}
    \frac{d \omega}{d t}=& \frac{1}{e}\sqrt{\frac{p}{Gm}}
    \left[-\cos\phi\,\mathcal{R}+\frac{2+e\cos\phi }{1+e\cos\phi}\mathcal{S}\right.\nonumber\\
    &\left.-e\cot\iota\frac{\sin(\omega+\phi)}{1+e\cos\phi}\mathcal{W} \right],\\
    \label{pk-ft}
    \frac{d \phi}{d t}=& \sqrt{\frac{G m}{p^3}}\left(1+e\cos\phi\right)^2\nonumber\\
    &+\frac{1}{e}\sqrt{\frac{p}{G m}}\left[\cos\phi\,\mathcal{R}-\frac{2+e\cos\phi}{1+e\cos\phi}\sin\phi\,\mathcal{S}\right].
\end{align}
Expanding the above equations in first-order approximation of $p^2/{Gm}$, we get
\begin{align}
    \label{pk-pf}
    \frac{d p}{d \phi}=& 2\frac{p^3}{Gm}\frac{1}{(1+e\cos\phi)^3}\mathcal{S},\\
    \label{pk-ef}
    \frac{d e}{d \phi}=& \frac{p^2}{Gm}\left[\frac{\sin\phi}{(1+e\cos\phi)^2}\mathcal{R}\right.\nonumber\\
    &\left.+\frac{2\cos\phi+e(1+\cos^2 \phi)}{(1+e\cos\phi)^3}\mathcal{S}\right],\\
    \label{pk-wf}
    \frac{d \omega}{d \phi}=& \frac{1}{e}\frac{p^2}{Gm}\left[-\frac{\cos\phi}{(1+e\cos\phi)^2}\mathcal{R}+\frac{(2+e\cos\phi){\sin\phi} }{(1+e\cos\phi)^3}\mathcal{S}\right.\nonumber\\
    &\left.-e\cot\iota\frac{\sin(\omega+\phi)}{(1+e\cos\phi)^3}\mathcal{W} \right],\\
    \label{pk-tf}
    \frac{d t}{d \phi}=& \sqrt{\frac{p^3}{G m}}\frac{1}{(1+e\cos\phi)^2}\left\{1-\frac{1}{e}\frac{p^2}{G m}\right.\nonumber\\
    &\left.\times\left[ \frac{\cos\phi}{(1+e\cos\phi)^2}\mathcal{R}-\frac{(2+e\cos\phi){\sin\phi} }{(1+e\cos\phi)^3}\mathcal{S}\right] \right\}.
\end{align}

In most applications, the orbital elements have two types of behaviors, i.e., the oscillatory and accumulated changes.
The oscillatory changes are oscillations with a period equal to one or multiple orbital period and can be averaged out after one or several cycles. 
The accumulated changes are steady drifts and cannot be averaged out with a few orbital cycles. 
The accumulated change of an arbitrary orbital element $K$ over a complete orbit is
\begin{equation}\label{accmulated}
    \Delta K=\int_{0}^{P}\frac{d K}{dt}\,dt=\int_{0}^{2\pi} \frac{d K}{d \phi} \,d\phi, 
\end{equation}
where $P$ is the orbital period. 
The orbital average of $dK/{d\phi}$ can be defined as
\begin{equation}\label{average}
  \left<\frac{d K}{d\phi}\right>=\frac{\Delta K}{P}=\frac{1}{P}\int_{0}^{P}\frac{d K}{dt}\,dt=\frac{1}{2\pi}\int_{0}^{2\pi} \frac{d K}{d \phi} \,d\phi.
\end{equation}

\section{Parameter estimation for IMRIs with DM minispikes in circular orbits}
\label{APPENDIX2}

In this section we discuss the parameter estimation with the FIM method.
To derive analytical GW waveforms, we consider IMRIs with DM minispikes in circular orbits
and ignore the change of small BH's mass.
Under the stationary phase approximation \cite{Cutler:1994ys,Will:1994fb}, 
the frequency-domain GW waveform in the inspiral stage is 
\begin{eqnarray}\label{h-ft}
    h(f)\simeq
        \begin{cases}
    \mathcal{A}f^{-7/6} e^{i\Psi(f)} ,    & 0< f< f_{\text{max}}, \\
    0,  & f> f_{\text{max}},
        \end{cases}
    \end{eqnarray}
where $\Psi(f)$ is the phase, the cutoff frequency $f_\text{max}$ is taken to be the frequency at the ISCO,
\begin{equation}
    f_\text{max}=\left(6^{3/2}\pi G m/c^3\right)^{-1},
\end{equation}
the amplitude $\mathcal{A}$ is
\begin{equation}
    \mathcal{A}=\frac{1}{\sqrt{30}\pi^{2/3}}\frac{(G\mathcal{M}_c)^{5/6}}{c^{2/3} d_L} \mathcal{X}(f),
\end{equation}
the chirp mass $\mathcal{M}_c=\eta^{3/2}m$, $d_L$ is the luminosity distance from the source to the observer, $\mathcal{X}(f)$ is defined as
\begin{equation}
    \mathcal{X}(f)=\left[ \frac{5}{96} \pi^{-8/3} G^{-5/3} \mathcal{M}^{-5/3}c^{5} f^{(2\alpha-11)/3}\,\kappa +1\right]^{-\frac{1}{2}},
\end{equation}
where
\begin{equation}
    \label{kappa}
    \kappa =\frac{12G\,\mu\,\rho_\text{sp}\,r_\text{sp}^{\alpha }(I_v+\lambda)}{m}\left(\frac{\pi ^2}{G m}\right)^{\alpha /3}.
\end{equation}
The phase $\Psi(f)$ is
\begin{equation}\label{gw-phase}
    \Psi(f)=2 \pi\, f\, t(f)-\frac{\pi }{4}-\psi(f),
\end{equation}
where $t(f)$ and $\psi(f)$ are
\begin{align}
    \label{t0-int}
    t(f)=& \int_{f_c}^{f}\left(\kappa f'^{2\alpha/3}+\frac{96\pi^{8/3}G^{5/3}\mathcal{M}_c^{5/3}f'^{11/3}}{5c^5}\right)^{-1}df',\\
    \label{phi0-int}
    \psi(f)=& 2\pi\int_{f_c}^{f}\left(\kappa
    f'^{2\alpha/3-1}\right.\nonumber\\
    &\left.+\frac{96\pi^{8/3}G^{5/3}\mathcal{M}_c^{5/3}f'^{8/3}}{5c^5}\right)^{-1}df',
\end{align}
$f_c=f_\text{max}$ is the frequency at the time of coalescence.
Introducing the variables
\begin{align}
    \label{Ff-int}
    F(f)&=\int \left(\frac{96 \pi ^{8/3} f^{8/3} G^{5/3} \mathcal{M}_c^{5/3}}{5 c^5}+\kappa f^{\frac{2 \alpha }{3}-1}\right)^{-1}  df,\\
    \label{Gf-int}
    G(f)&=\int \left(\frac{96 \pi ^{8/3} f^{11/3} G^{5/3} \mathcal{M}_c^{5/3}}{5 c^5}+\kappa f^{\frac{2 \alpha }{3}}\right)^{-1} df,
\end{align}
and substituting Eqs. \eqref{t0-int}, \eqref{phi0-int}, \eqref{kappa}, 
\eqref{Ff-int}, and \eqref{Gf-int} into Eq. \eqref{gw-phase}, the phase can be written as
\begin{equation}
    \begin{split}
    \Psi(f)
    =2\,\pi\,f\,t_c+2\pi[f\,G(f)-F(f)]-\Phi_c-\frac{\pi}{4},
    \end{split}
\end{equation}
where $t_c=-G(f_c)$ and $\Phi_c=-F(f_c)$ are the time and phase at the coalescence, respectively.

With the GW waveform \eqref{h-ft}, we calculate the FIM,
\begin{equation}\label{FIM}
    \Gamma_{ab}=\left.\left(\frac{\partial h}{\partial \theta^a} \right | \frac{\partial h}{\partial \theta^b} \right),
    \end{equation}
where $\theta^a=\{\phi_c,t_c,\ln \mathcal{M}_c,\ln\kappa,\alpha\}$ are source's parameters.
The estimated error of the parameter $\theta^a$ in the large SNR limit is
\begin{equation}
    \label{FIM-rms}
    \Delta\theta^a=\sqrt{\Sigma_{aa}},
\end{equation}
where the inverse of the FIM is $\Sigma_{ab}=(\Gamma^{-1})_{ab}$.
The corresponding partial derivatives of $\tilde{h}(f)$ are
\begin{align}
    \label{pd-phic}
    \frac{\partial \tilde{h}(f)}{\partial \Phi _c}=& -i \tilde{h}(f),\\
    \label{pd-tc}
    \frac{\partial\tilde{h}(f)}{\partial t_c}=& 2\pi i\tilde{h}(f),\\
    \label{pd-cm}
    \frac{\partial\,\tilde{h}(f)}{\partial \ln\mathcal{M}_c}=& \mathcal{M}_c\left[\frac{\partial \mathcal{A}}{\partial \mathcal{M}_c}f^{-7/6}e^{i\Psi(f)}\right.\nonumber\\
    &\left.+i \tilde{h}(f)\left(2\pi f \frac{\partial G(f)}{\partial \mathcal{M}_c}-\frac{\partial F(f)}{\partial \mathcal{M}_c}\right)\right],\\
    \label{pd-ka}
    \frac{\partial\tilde{h}(f)}{\partial\ln\kappa}=&\kappa\bigg[\frac{\partial \mathcal{A}}{\partial \kappa}f^{-7/6}e^{i\Psi(f)}\nonumber\\
    &+i\,\tilde{h}(f)\left(2 \pi f \frac{\partial G(f)}{\partial \kappa }-\frac{\partial F(f)}{\partial \kappa }\right)\bigg],\\
    \label{pd-alpha}
    \frac{\partial\tilde{h}(f)}{\partial \alpha}=&\frac{\partial \mathcal{A}}{\partial \alpha}f^{-7/6}e^{i\Psi(f)}\nonumber\\
    &+ i \tilde{h}(f)\left(2 \pi f \frac{\partial G(f)}{\partial \alpha }-\frac{\partial F(f)}{\partial \alpha }\right).
\end{align}

Since the source comes from all directions, we use the averaged response function. The effective noise PSD is
\begin{equation}
\label{LISASn}
    S_n(f)=\frac{S_h(f)}{R_n(f)},
\end{equation}
where the analytical expression for the sky and polarization averaged response function  $R_n(f)$ of spaced-based GW detectors was derived in \cite{Zhang:2020khm}.
For simplicity,
we take $\Phi_c=0$, $t_c=0$, and four-year observation time prior to ISCO with SNR$=10$ \cite{Will:1994fb,Yagi:2009zm}. 
For LISA, the lower and upper cutoff frequencies are $f_\text{low}=10^{-5}$ Hz and $f_\text{high}=1$ Hz \cite{Yagi:2009zm},
so the lower and upper limits of the integration in Eq. \eqref{overlap} are chosen as $f_\text{ini}=\text{Max}(f_\text{low}, f_\text{4yr})$ and $f_\text{end}=\text{Min}(f_\text{high}, f_\text{ISCO})$, respectively. Here $f_\text{4yr}$ is the frequency at four years before the ISCO. 
We then use Eq. \eqref{FIM-rms} to estimate the parameter errors for $10M_\odot/10^3M_\odot$ IMRIs with $\alpha=2.25$, $7/3$, and $2.5$.

%\bibliographystyle{ref01}
%\bibliography{DM-ref}


\end{document}